\documentclass[aip,jcp,reprint]{revtex4-2}

\usepackage{amsmath}
\usepackage{upgreek}
\usepackage{float}
\usepackage{graphicx}
\usepackage{tabularx}
\usepackage{hyperref}

\begin{document}

\title{Insights into lithium manganese oxide-water interfaces using machine learning potentials}

\author{Marco Eckhoff}
\email{marco.eckhoff@chemie.uni-goettingen.de}
\affiliation{Universit\"at G\"ottingen, Institut f\"ur Physikalische Chemie, Theoretische Chemie, Tammannstra{\ss}e 6, 37077 G\"ottingen, Germany.}
\author{J\"org Behler}
\email{joerg.behler@uni-goettingen.de}
\affiliation{Universit\"at G\"ottingen, Institut f\"ur Physikalische Chemie, Theoretische Chemie, Tammannstra{\ss}e 6, 37077 G\"ottingen, Germany.}
\affiliation{Universit\"at G\"ottingen, International Center for Advanced Studies of Energy Conversion (ICASEC), Tammannstra{\ss}e 6, 37077 G\"ottingen, Germany.}

\date{\today}

\begin{abstract}
Unraveling the atomistic and the electronic structure of solid-liquid interfaces is the key to the design of new materials for many important applications, from heterogeneous catalysis to battery technology. Density functional theory (DFT) calculations can in principle provide a reliable description of such interfaces, but the high computational costs severely restrict the accessible time and length scales. Here, we report machine learning-driven simulations of various interfaces between water and lithium manganese oxide (Li$_x$Mn$_2$O$_4$), an important electrode material in lithium ion batteries and a catalyst for the oxygen evolution reaction. We employ a high-dimensional neural network potential (HDNNP) to compute the energies and forces several orders of magnitude faster than DFT without loss in accuracy. In addition, a high-dimensional neural network for spin prediction (HDNNS) is utilized to analyze the electronic structure of the manganese ions. Combining these methods, a series of interfaces is investigated by large-scale molecular dynamics. The simulations allow us to gain insights into a variety of properties like the dissociation of water molecules, proton transfer processes, and hydrogen bonds, as well as the geometric and electronic structure of the solid surfaces including the manganese oxidation state distribution, Jahn-Teller distortions, and electron hopping.
\end{abstract}

\keywords{Machine Learning Potentials, High-Dimensional Neural Networks, Molecular Dynamics Simulations, PBE0r Local Hybrid Density Functional, Lithium Manganese Oxide-Water Interface, Oxidation States, Electron Hopping, Interfacial Water, Water Dissociation, Hydroxide Layer, Proton Transfer}

\maketitle

%%%%%%%%%%%%%%%%%%%%%%%%%%%%%%%%%%%%%%%%%%%%%%%%%%%%%%%%%%%%%%%%%%%%%%%%%%%%%%%%%%%%%%%%%%%%%%%%%%%%
\section{Introduction}
%%%%%%%%%%%%%%%%%%%%%%%%%%%%%%%%%%%%%%%%%%%%%%%%%%%%%%%%%%%%%%%%%%%%%%%%%%%%%%%%%%%%%%%%%%%%%%%%%%%%

The understanding of solid-liquid interfaces is of major importance for a sustainable energy future.\cite{Chu2012, Larcher2015} In particular, electrode-electrolyte interfaces are central for many processes, from the electrocatalytic water splitting for the production of green hydrogen to energy storage in lithium ion batteries supplying, e.g., portable electronic devices and electric vehicles.\cite{Bruce2008, Goodenough2010, Suntivich2011, Seh2017, Suen2017} For this purpose, a prominent material is the lithium manganese oxide spinel Li$_x$Mn$_2$O$_4$, with $0\leq x\leq2$, which is a frequently used positive electrode material in lithium ion batteries but can also be employed as electrocatalyst for the oxygen evolution reaction (OER) representing the limiting step of water splitting.\cite{Thackeray1983, Thackeray1997, Cady2015, Koehler2017} Beyond conventional lithium ion batteries containing organic electrolytes, the Li$_x$Mn$_2$O$_4$-water interface has recently received increasing attention regarding the development of environment-friendly aqueous rechargeable lithium ion batteries offering improved safety combined with higher ionic conductivity and lower-cost production.\cite{Kim2014, Alias2015}

Stoichiometric LiMn$_2$O$_4$ contains a one-to-one ratio of Mn$^\mathrm{III}$ and Mn$^\mathrm{IV}$ ions.\cite{Takahashi2003, Akimoto2004, Piszora2004} The electrochemical incorporation or removal of Li ions during battery discharging and charging changes this ratio by the reduction or oxidation of Mn ions, respectively, ensuring overall charge neutrality\cite{Hunter1981, Thackeray1983} and allowing to control the electronic structure of the bulk material. This control is particularly important for the OER, as the Mn oxidation states are considered to be central for this process.\cite{Stamenkovic2006} Apart from the overall composition of the material, the OER activity is also determined by the details of the solid-electrolyte interface whose geometric and electronic structure as well as the atomic composition can be substantially different from the bulk.\cite{Gauthier2015, Schoenewald2020}

The active sites of electrocatalytic reactions are often embedded in a complex environment consisting of the -- possibly reconstructed -- solid surface and the electrical double layer. Small particles and porous materials have large surface-to-volume ratios, which are beneficial for a high activity, but the exposed surfaces can exhibit very different reactivities. The identification of active sites is therefore essential in a bottom-up approach for the design of improved catalysts.\cite{Hammer2000, Jaramillo2007, Norskov2009, Behrens2012, Jiao2015}

A practical challenge when using Li$_x$Mn$_2$O$_4$ as battery material is capacity fading, which is related to the disproportionation of Mn$^\mathrm{III}$ ions at the interface and the dissolution of the resulting Mn$^\mathrm{II}$ ions.\cite{Benedek2012, Bhandari2017, Leung2017} Consequently, the identification of tactics for controlling the Mn oxidation states at the interface is important for the construction of batteries with improved charge/discharge cycles and enhanced lifetime. Hence, to unravel the relationship between composition and reactivity, a comprehensive understanding of the geometric and electronic structure as well as of the dynamics and reactions at the interface is required.\cite{Hirayama2010, Leung2013, Benedek2017}

These insights can be gained in principle in computer simulations, but complex interface systems still pose a significant challenge as they require a first principles-quality description.\cite{Schaub2001, Rossmeisl2007, Bajdich2013, Tocci2014, Friebel2015} The electronic structure of Li$_x$Mn$_2$O$_4$ with coexisting Mn$^\mathrm{IV}$ and Jahn-Teller distorted high-spin Mn$^\mathrm{III}$ ions in the bulk as well as high-spin Mn$^\mathrm{II}$ ions at the interface is, however, difficult to describe by established methods like density functional theory (DFT). For a correct representation at least the level of the generalized gradient approximation including an additional Hubbard-like term for on-site Coulomb interactions (GGA$+U$) or a hybrid functional containing a fraction of exact Hartree-Fock exchange is needed.\cite{Karim2013, Kumar2014, Lee2016, Warburton2016, Eckhoff2020} A recent hybrid DFT benchmark of lithium manganese oxides showed that on-site Hartree-Fock exchange terms yield a correct description of partially filled shells of localized d electrons.\cite{Eckhoff2020} However, up to now ab initio molecular dynamics simulations using GGA$+U$ or hybrid DFT functionals could only be performed for rather small Li$_x$Mn$_2$O$_4$-water model systems containing a few hundred atoms on picosecond time scales due to the large computational effort.\cite{Leung2012, Choi2018, Intan2019, Okuno2019, Zhou2020} To consider the interplay of a variety of different structural motifs with a liquid solvent, picosecond time scales are not sufficient. For instance, for the equilibration of Li$_x$Mn$_2$O$_4$-water interfaces including the formation of hydroxide layers, electrical double layers, and/or strongly bound water at the interface as well as to obtain reliable statistics for elementary steps of proton transfer (PT) reactions and hydrogen bond networks significantly larger length and time scales are required.

Machine learning potentials (MLP) combine the efficiency of simple empirical potentials with the accuracy of quantum mechanics allowing to meet these requirements.\cite{Behler2016, Bartok2017, Noe2020, Behler2021a} Consequently, various MLPs have been developed, e.g., for water\cite{Morawietz2016, Cheng2016, Zhang2018, Cheng2019} and different solid-liquid interface systems.\cite{Wen2019, Artrith2019, Pun2020, Schran2021} In previous works we could further show that a high-dimensional neural network potential (HDNNP),\cite{Behler2007, Behler2014, Behler2015, Behler2017, Behler2021} a frequently used type of MLPs, is applicable to bulk materials containing transition metal ions in different oxidation states\cite{Eckhoff2020a, Eckhoff2020b, Eckhoff2021} and in different magnetic orders.\cite{Eckhoff2021} The ability to represent different oxidation states is mandatory for studying the Li$_x$Mn$_2$O$_4$ system. As the HDNNP's underlying functional form is unbiased with respect to different interaction types, an equally reliable description for the interactions in the bulk and at the interface is obtained as has been demonstrated, e.g., for the Cu-water\cite{Natarajan2016, Natarajan2017} and ZnO-water interfaces.\cite{Quaranta2017, Hellstroem2019, Quaranta2019} Moreover, like most MLPs, HDNNPs are reactive, i.e., they are able to describe the formation and cleavage of bonds making them applicable to PT reactions omnipresent in electrochemical systems.

Apart from the simulation of the atomistic structure, machine learning algorithms can also be applied to obtain information about the electronic structure, for example, atomic charges,\cite{Artrith2011, Ghasemi2015, Unke2019, Xie2020, Ko2020} electrostatic multipole moments,\cite{Houlding2007, Bereau2015} polarizabilities,\cite{Montavon2013} and even quantum mechanical wavefunctions.\cite{Schuett2019} Moreover, our recently developed high-dimensional neural network spin (HDNNS) prediction method\cite{Eckhoff2020b} can be used to identify the oxidation and spin states of the Mn ions. This method is based on local geometric changes associated to the details of the electronic structure. Thus, in combination with an HDNNP providing the energies and forces, nowadays nanosecond time scale investigations of the geometric and electronic structure are possible for systems containing about $10^4$-$10^5$ atoms.

In this work we investigate the Li$_x$Mn$_2$O$_4$-water interface employing an HDNNP in combination with an HDNNS. Several \{100\} and \{110\} surfaces with different terminations in contact with water are investigated to unravel the spatial distribution of Mn oxidation states and oxygen species, such as oxide ions O$^{2-}$, hydroxide ions OH$^-$, neutral water molecules H$_2$O, and hydronium ions H$_3$O$^+$. Starting from atomically smooth solid surfaces in contact with a liquid water phase, the formation of hydroxide and strongly bound water layers at the interface is studied to understand the fundamental properties of the interface structure. A detailed analysis of Jahn-Teller distortions and the hydrogen bond network is provided in the Supplementary Material. The analysis of electron hopping rates between Mn ions, various PT reactions, and water species residence lifetimes under equilibrium conditions yields a detailed understanding of the kinetics and dynamics. Finally, we compare the activity of the different surfaces and sites and show that initial steps of the OER occur already spontaneously under equilibrium conditions.

%%%%%%%%%%%%%%%%%%%%%%%%%%%%%%%%%%%%%%%%%%%%%%%%%%%%%%%%%%%%%%%%%%%%%%%%%%%%%%%%%%%%%%%%%%%%%%%%%%%%
\section{Methods}
%%%%%%%%%%%%%%%%%%%%%%%%%%%%%%%%%%%%%%%%%%%%%%%%%%%%%%%%%%%%%%%%%%%%%%%%%%%%%%%%%%%%%%%%%%%%%%%%%%%%

In HDNNPs\cite{Behler2007, Behler2021}, which we use to compute the energies and forces driving the molecular dynamics (MD) simulations, the potential energy is constructed as a sum of atomic energy contributions $E_n^\alpha$, 
\begin{align}
E(\{\mathbf{R}\})=\sum_{\alpha=1}^{N_\mathrm{elem}}\sum_{n=1}^{N_\mathrm{atoms}^{\alpha}}E_n^\alpha\ .
\end{align}
Here, $\{\mathbf{R}\}$ are the nuclear coordinates for a system containing $N_\mathrm{elem}$ elements with $N_\mathrm{atoms}^{\alpha}$ for element $\alpha$. 
The individual atomic energy contributions are represented by atomic feed-forward neural networks of the form
\begin{align}
\begin{split}
E_n=&\ b_1^4+\sum_{l=1}^{n_3}a_{l1}^{34}\cdot\tanh\Bigg\{b_l^3+\sum_{k=1}^{n_2}a_{kl}^{23}\cdot\tanh\Bigg[b_k^2\\
&\,+\sum_{j=1}^{n_1}a_{jk}^{12}\cdot\tanh\Bigg(b_j^1+\sum_{i=1}^{n_G}a_{ij}^{01}\cdot G_{n,i}\Bigg)\Bigg]\Bigg\}\ .
\end{split}\label{eq:energy_contributions}
\end{align}

The architecture $n_G$-$n_1$-$n_2$-$n_3$-1 of the atomic neural networks contains an input layer with $n_G$ neurons providing a description of the atomic environment. Moreover, in this work three hidden layers with $n_1$, $n_2$, and $n_3$ neurons, respectively, and an output layer with one neuron, which yields the atomic energy contribution, are used. The activation functions are hyperbolic tangents except for the output layer, for which a linear function is employed. The weight parameters $\{a_{\mu\nu}^{\rho\sigma}\}$ and $\{b_\nu^\sigma\}$ of the atomic neural networks are optimized to accurately reproduce a training data set consisting of energies $E^\mathrm{ref}$ and atomic force components $F^\mathrm{ref}$ of reference structures obtained, for example, in DFT calculations. For each element an individual atomic neural network is constructed making the neural network parameters element-specific. For clarity, the index $\alpha$ representing this element-dependence has been omitted in the quantities in Equation \ref{eq:energy_contributions}.

The atomic neural networks are able to describe the complex relation between the atomic energies and the local chemical environments of the atoms. These environments are described by vectors of many-body atom-centered symmetry functions (ACSF)\cite{Behler2011} $\mathbf{G}_n^\alpha$ serving as structural fingerprints of the local geometry inside a cutoff sphere of radius $R_\mathrm{c}$. ACSFs represent a general transformation from the Cartesian coordinates $\{\mathbf{R}\}$ to a translationally, rotationally, and permutationally invariant structural description based on interatomic distances and angles. Moreover, for all atoms of a given element, the ACSF vectors have the same dimensionality to ensure the applicability of the trained atomic neural networks to large-scale simulations of systems containing different numbers of atoms. As the ACSFs depend only on the elements and positions of the atoms, HDNNPs are able to describe the making and breaking of bonds. The parameters defining the spatial shapes of the radial and angular ACSFs can be adjusted to optimize the performance as described in the Supplementary Material. More detailed information about HDNNPs, ACSFs, their properties, and their construction are provided in several reviews.\cite{Behler2014, Behler2015, Behler2017, Behler2021}

The HDNNS method\cite{Eckhoff2020b} is closely related to HDNNPs and employs the same atomic neural network topology. However, instead of atomic energy contributions $E_n^{\alpha}$, the atomic neural networks (Equation \ref{eq:energy_contributions}) yield the atomic spins $S_n^{\alpha}$, i.e., they provide the $S_n^\alpha(\{\mathbf{R}\})$ relation. A HDNNS exploits the observation that different oxidation states as well as high- and low-spin states of transition metal ions typically lead to structurally different local environments. Thus, like for the energies and forces in HDNNPs, the method is based on the assumption that the atomic spins and oxidation states are uniquely defined by the structure. Consequently, consistent reference data corresponding to the ground state electronic structure are mandatory for a successful construction of the HDNNS. The absolute values of atomic reference spins obtained from DFT are used for training to circumvent the issue that the electronic ground state is twofold degenerate with respect to the absolute sign of all spins. Since the present work is restricted to the magnetic ground state, we do not explicitly include the degrees of freedom related to the relative orientations of the atomic spins, but we note that magnetic HDNNPs taking these degrees of freedom into account have been proposed.\cite{Eckhoff2021}

Using both, an HDNNP and an HDNNS, in MD simulations enables a simultaneous first principles-quality representation of the geometric and the qualitative electronic structure dynamics on nanosecond time scales for systems containing thousands of atoms.\cite{Eckhoff2020a, Eckhoff2020b, Eckhoff2021}

%%%%%%%%%%%%%%%%%%%%%%%%%%%%%%%%%%%%%%%%%%%%%%%%%%%%%%%%%%%%%%%%%%%%%%%%%%%%%%%%%%%%%%%%%%%%%%%%%%%%
\section{Computational Details}
%%%%%%%%%%%%%%%%%%%%%%%%%%%%%%%%%%%%%%%%%%%%%%%%%%%%%%%%%%%%%%%%%%%%%%%%%%%%%%%%%%%%%%%%%%%%%%%%%%%%

For the generation of the reference data the local hybrid exchange-correlation functional PBE0r\cite{Sotoudeh2017, Eckhoff2020} including D3 dispersion corrections\cite{Grimme2010, Grimme2011} was used in collinar spin-polarized DFT calculations. PBE0r considers only on-site Hartree-Fock exchange terms yielding an accurate description of the partially filled Mn d shell with a computational effort comparable to generalized gradient approximation functionals. The Car-Parrinello Projector Augmented-Wave (CP-PAW) code (version from September 28, 2016)\cite{Bloechl1994, CP-PAW} and the DFT-D3 code (version from June 14, 2016)\cite{Grimme2010, Grimme2011} were employed using the same setup as in our previous studies.\cite{Eckhoff2020, Eckhoff2020a, Eckhoff2020b}

The HDNNP and HDNNS were constructed using the RuNNer code (versions from October 19, 2020 and December 4, 2018, respectively).\cite{Behler2015, Behler2017, RuNNer} The architecture of the atomic neural networks is 180-25-20-15-1 for all elements in the HDNNP and 180-20-15-10-1 for Mn in the HDNNS, which is the only spin-polarized atom in the system. The parameters of the 180 radial and angular ACSFs per element with the cutoff radius $R_\mathrm{c}=10.5\,a_0$ are compiled in the Supplementary Material along with the description of a generally applicable scheme to adjust the parameter $\eta$ of the ACSFs to the element-specific nearest-neighbor distances. $a_0$ is the Bohr radius. 

Instead of total energies, the DFT formation energies were used for training, which were obtained from the total energies minus the sum of the atomic energies calculated for the elements in their reference states, i.e., gaseous H$_2$, body centered cubic Li, gaseous O$_2$, and $\upalpha$-Mn. In addition to the formation energies also the DFT Cartesian atomic force components were used for training the HDNNP. Of all available energies and force components 90{\%} were used in the training set to determine the neural network parameters, while the remaining 10{\%} were employed to test the predictive power and reliability for structures not included in the training process. 

The DFT reference atomic spins used to train the HDNNS are the absolute values of projections of the spin density onto the one-center expansions of the partial waves using atom-centered spheres with a radius of 1.2 times the atomic covalent radius. The atomic spin is therefore equal to the absolute difference in the number of spin-up and spin-down electrons at an atom in units of the electron spin $\tfrac{1}{2}\,\hbar$. The setup of the HDNNP and HDNNS construction is described in detail in the Supplementary Material.

\begin{figure*}[htb!]
\centering
\includegraphics[width=\textwidth]{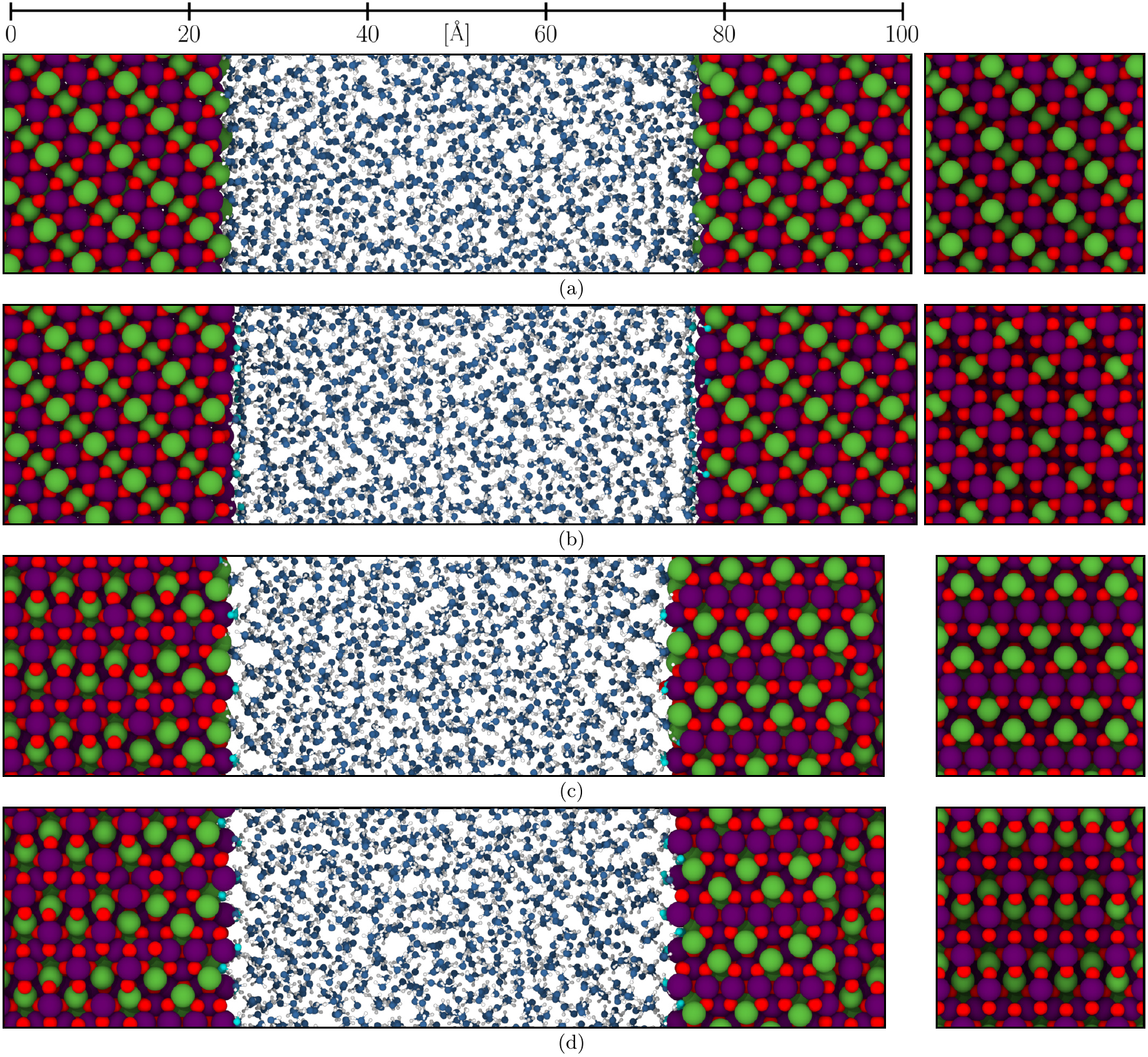}
\caption{Side views ($zy$ plane) of equilibrated Li$_x$Mn$_2$O$_4$-water interface simulation cells with a volume of about $25\cdot25\cdot100\,\text{\AA}^3$ including the interfaces (a) $\{100\}_\mathrm{Li}$, (b) $\{100\}_{\mathrm{Mn}_2\mathrm{O}_4}$, (c) $\{110\}_{\mathrm{LiMnO}_2}$, and (d) $\{110\}_{\mathrm{MnO}_2}$ are shown on the left. Top views ($xy$ plane) of non-equilibrated vacuum solid surfaces are provided on the right. The scale bar applies to all panels. The atoms are shown as balls whereby Li is colored green, Mn violet, and H white. The color/size of the oxygen atoms has been chosen according to the H connectivity: O$^{2-}$ is red/large, OH$^-$ turquoise/small, H$_2$O blue/small, and H$_3$O$^+$ orange/small. The O-H bonds are shown as sticks. The surrounding black line corresponds to the boundaries of the periodic simulation cell. A side view on the $zx$ plane is provided in the Supplementary Material Figures S1 (a) to (d). This figure was created with OVITO Pro (version 3.5.0).\cite{Ovito}}\label{fig:structure}
\end{figure*}

HDNNP-driven MD simulations were performed for Li$_x$Mn$_2$O$_4$-water interface systems in the isothermal-isobaric ($NpT$) ensemble at a temperature of $T=298$\,K and a pressure of $p=1$\,bar. The simulation cells include about $N\approx6\cdot10^3$ atoms in a volume of about $25\cdot25\cdot100\,\text{\AA}^3$. The volume ratio of Li$_x$Mn$_2$O$_4$ and water is approximately 1:1 with the phase boundaries parallel to the $xy$ plane. Four different cuts of bulk Li$_x$Mn$_2$O$_4$ in contact with water have been investigated, which are the \{100\} Li-terminated surface with Mn$_2$O$_4$ in the first subsurface layer ($\{100\}_\mathrm{Li}$), the \{100\} Mn$_2$O$_4$-terminated surface with Li in the first subsurface layer ($\{100\}_{\mathrm{Mn}_2\mathrm{O}_4}$), the \{110\} LiMnO$_2$-terminated surface ($\{110\}_{\mathrm{LiMnO}_2}$), and the \{110\} MnO$_2$-terminated surface ($\{110\}_{\mathrm{MnO}_2}$). Top views of the clean surfaces as well as side views of the employed interface slab models are shown in Figures \ref{fig:structure} (a) to (d), respectively. Both surfaces of each slab are structurally identical. Therefore, the solid slab is built from bulk stoichiometric LiMn$_2$O$_4$ with non-stoichiometric Li$_x$Mn$_2$O$_4$ surfaces. For each system, three simulations starting from different H$_2$O configurations, which initially did not include OH$^-$ and H$_3$O$^+$ ions, on top of the atomically flat solid surfaces were performed for an equilibration time of 1\,ns and a subsequent acquisition time of 5\,ns. In the same way, three bulk water simulations with different initial structures were carried out using approximately cubic cells containing around $1.5\cdot10^4$ atoms and fluctuating lattice parameters between about 50 and 60\,{\AA} in the $NpT$ ensemble. To identify the oxygen and water species, each H atom has been assigned to its closest O atom.

The HDNNP-driven simulations were performed using the Large-scale Atomic/Molecular Massively Parallel Simulator (LAMMPS)\cite{Plimpton1995, LAMMPS} and the neural network potential package (n2p2).\cite{Singraber2019, n2p2} They were run with a timestep of 0.5\,fs applying the Nos\'{e}-Hoover thermostat and barostat\cite{Nose1984, Hoover1985} with coupling constants of 0.05\,ps and 0.5\,ps, respectively, allowing for anisotropic changes of the simulation cell. The trajectory was stored in intervals of 0.1\,ps.

%%%%%%%%%%%%%%%%%%%%%%%%%%%%%%%%%%%%%%%%%%%%%%%%%%%%%%%%%%%%%%%%%%%%%%%%%%%%%%%%%%%%%%%%%%%%%%%%%%%%
\section{Results and Discussion}
%%%%%%%%%%%%%%%%%%%%%%%%%%%%%%%%%%%%%%%%%%%%%%%%%%%%%%%%%%%%%%%%%%%%%%%%%%%%%%%%%%%%%%%%%%%%%%%%%%%%

%%%%%%%%%%%%%%%%%%%%%%%%%%%%%%%%%%%%%%%%%%%%%%%%%%%%%%%%%%%%%%%%%%%%%%%%%%%%%%%%%%%%%%%%%%%%%%%%%%%%
\subsection{High-dimensional neural networks}
%%%%%%%%%%%%%%%%%%%%%%%%%%%%%%%%%%%%%%%%%%%%%%%%%%%%%%%%%%%%%%%%%%%%%%%%%%%%%%%%%%%%%%%%%%%%%%%%%%%%

The HDNNP and HDNNS are based on a reference data set consisting of 15228 Li$_x$Mn$_2$O$_4$ bulk structures,\cite{Eckhoff2020a, Eckhoff2020b} 5143 water bulk structures, and 17597 Li$_x$Mn$_2$O$_4$-water interface structures and their PBE0r-D3 DFT energies, atomic force components, and atomic spins. The structures include 32 to 255 atoms, with the interface structures containing between 122 and 194 atoms. A detailed description of the reference data set construction and composition is provided in the Supplementary Material.

\begin{figure*}[htb!]
\centering
\includegraphics[width=\textwidth]{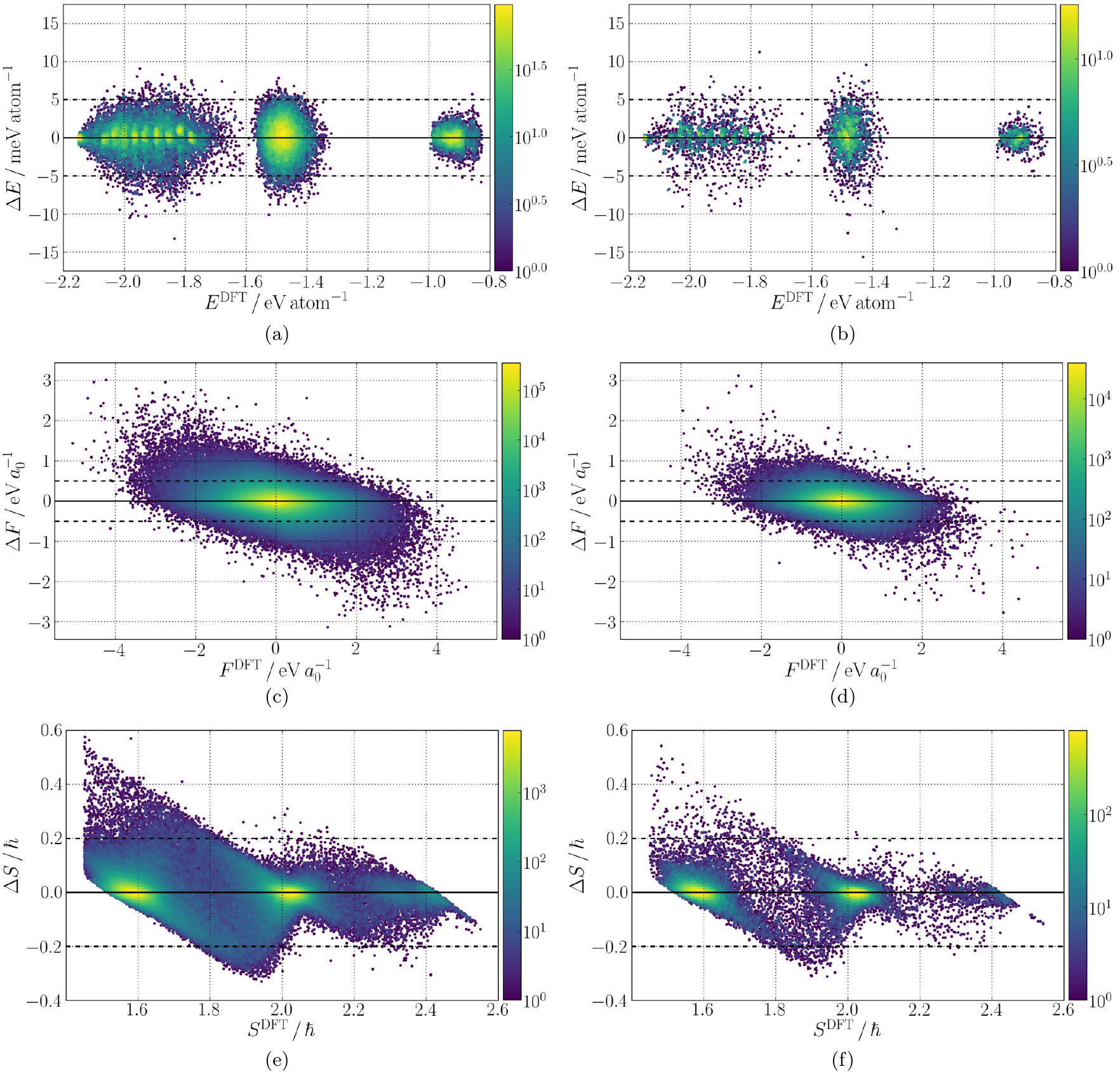}
\caption{Energy errors $\Delta E=E^\mathrm{HDNNP}-E^\mathrm{DFT}$ as a function of the reference formation energy $E^\mathrm{DFT}$ of the (a) training set and (b) test set, force component errors $\Delta F=F^\mathrm{HDNNP}-F^\mathrm{DFT}$ as a function of the reference force components $F^\mathrm{DFT}$ of the (c) training set and (d) test set, and errors of the atomic spins $\Delta S=S^\mathrm{HDNNP}-S^\mathrm{DFT}$ as a function of the reference atomic spins $S^\mathrm{DFT}$ of the (e) training set and (f) test set. The color in the heatmaps represents the density of data points based on discretizing the plotting areas into grids of $200\times125$ points.}\label{fig:HDNN_quality}
\end{figure*}

The PBE0r-D3 DFT formation energies range from $-2.15$\,eV\,atom$^{-1}$ to $-0.83$\,eV\,atom$^{-1}$ therefore spanning a fitting interval of $1.32$\,eV\,atom$^{-1}$. In Figures \ref{fig:HDNN_quality} (a) and (b) the three structure types can be identified by their formation energies between about $-2.2$ and $-1.6$, $-1.6$ and $-1.3$, as well as $-1.0$ and $-0.8$\,eV\,atom$^{-1}$ corresponding to bulk Li$_x$Mn$_2$O$_4$, interface structures, and bulk water, respectively. 

The HDNNP reproduces the energies of the training set with a root mean squared error (RMSE) of 1.9\,meV\,atom$^{-1}$, while it is able to predict the energies of the test set with an RMSE of 2.4\,meV\,atom$^{-1}$. The RMSE of only the bulk water training structures is about 1.0\,meV\,atom$^{-1}$ (test set 1.1\,meV\,atom$^{-1}$) and thus about half of the values of the RMSEs of the Li$_x$Mn$_2$O$_4$ and interface training structures, which are 2.0 and 2.1\,meV\,atom$^{-1}$, respectively (test sets 2.3 and 2.8\,meV\,atom$^{-1}$). We ascribe this difference to the less complex geometric and electronic structure of bulk liquid water that is typically very well represented by machine learning potentials.\cite{Morawietz2016, Cheng2019} We note that in spite of the increased complexity of the potential energy surface of the interface system, in particular the bulk Li$_x$Mn$_2$O$_4$ test set RMSE is very similar to the RMSE of 2.2\,meV\,atom$^{-1}$ of a previous HDNNP fitted to bulk Li$_x$Mn$_2$O$_4$ data only.\cite{Eckhoff2020a} The maximum energy error of all data in the training/test set is 13.2/15.6\,meV\,atom$^{-1}$. Only 2.3/5.5{\%} of the energy predictions in the training/test set have errors larger than 5\,meV\,atom$^{-1}$ (Figures \ref{fig:HDNN_quality} (a) and (b)).

The PBE0r-D3 reference data contains force components up to $|F^\mathrm{DFT}|\leq 5.06$\,eV\,$a_0^{-1}$. The force components RMSE of all data is 0.127\,eV\,$a_0^{-1}$ for both, training and test set. Again, the RMSE of bulk water (0.059\,eV\,$a_0^{-1}$) is lower than the RMSEs of bulk Li$_x$Mn$_2$O$_4$ (0.114\,eV\,$a_0^{-1}$) and the interface structures (0.143\,eV\,$a_0^{-1}$). The force components RMSE of bulk Li$_x$Mn$_2$O$_4$ is similar to the value of 0.107\,eV\,$a_0^{-1}$ of the aforementioned bulk only Li$_x$Mn$_2$O$_4$ HDNNP.\cite{Eckhoff2020a} As highlighted by the heatmaps in Figures \ref{fig:HDNN_quality} (c) and (d) most of the force components have an error smaller than 0.5\,eV\,$a_0^{-1}$ (99.33/99.32{\%}). The maximum errors are 3.13 and 3.12\,eV\,$a_0^{-1}$ for the training and test set. 

The atomic spins of Mn are in the range $1.45\,\hbar\leq S^\mathrm{DFT}\leq 2.55\,\hbar$ (Figures \ref{fig:HDNN_quality} (e) and (f)). The RMSE of the HDNNS is $0.04\,\hbar$ for both training and test set. Only 0.75{\%} of the training data and 0.76{\%} of the test data show errors larger than $0.2\,\hbar$ possibly resulting in the assignment of a different spin and oxidation state, while the maximum errors are 0.57 and $0.54\,\hbar$, respectively. Consequently, the vast majority of the Mn oxidation states is accurately predicted. The assignment to the oxidation states Mn$^\mathrm{IV}$ (d electron configuration t$_\mathrm{2g}^3$e$_\mathrm{g}^0$), high-spin Mn$^\mathrm{III}$ (t$_\mathrm{2g}^3$e$_\mathrm{g}^1$), and high-spin Mn$^\mathrm{II}$ (t$_\mathrm{2g}^3$e$_\mathrm{g}^2$) has been set to the intervals $1.4\,\hbar\leq S<1.8\,\hbar$, $1.8\,\hbar\leq S<2.2\,\hbar$, and $2.2\,\hbar\leq S<2.6\,\hbar$, respectively, based on the distribution of spins shown in Figures \ref{fig:HDNN_quality} (e) and (f).

%%%%%%%%%%%%%%%%%%%%%%%%%%%%%%%%%%%%%%%%%%%%%%%%%%%%%%%%%%%%%%%%%%%%%%%%%%%%%%%%%%%%%%%%%%%%%%%%%%%%
\subsection{Manganese oxidation state distribution}\label{sec:Mn_oxidation_state}
%%%%%%%%%%%%%%%%%%%%%%%%%%%%%%%%%%%%%%%%%%%%%%%%%%%%%%%%%%%%%%%%%%%%%%%%%%%%%%%%%%%%%%%%%%%%%%%%%%%%

Using the obtained HDNNP and HDNNS, we investigate four solid-liquid interface systems with different geometric and electronic structure. In this section we first investigate the interface from the perspective of the solid phase, while we focus on the liquid phase in the next section and finally discuss the reactivity of interfacial water species in the last section.

In general, HDNNP-driven MD simulations allow to gain insights into the atomic structure and dynamics of the Li$_x$Mn$_2$O$_4$-water interface in equilibrium. Further, using the HDNNS we can in addition investigate the Mn oxidation state distribution in the trajectories. This distribution is of major importance for understanding capacity fading of Li$_x$Mn$_2$O$_4$ during battery usage because previous studies propose the origin to be disproportionation of Mn$^\mathrm{III}$ ions at the interface and subsequent dissolution of the emerging Mn$^\mathrm{II}$ ions. Mn$^\mathrm{II}$ is the most stable oxidation state in aqueous solution while Mn$^\mathrm{IV}$ is not soluble.\cite{Davies1969} Structural features exposing only small amounts of Mn$^\mathrm{II}$ and Mn$^\mathrm{III}$ ions at the interface could thus support the development of more durable electrode materials.

\begin{figure*}[htb!]
\centering
\includegraphics[width=\textwidth]{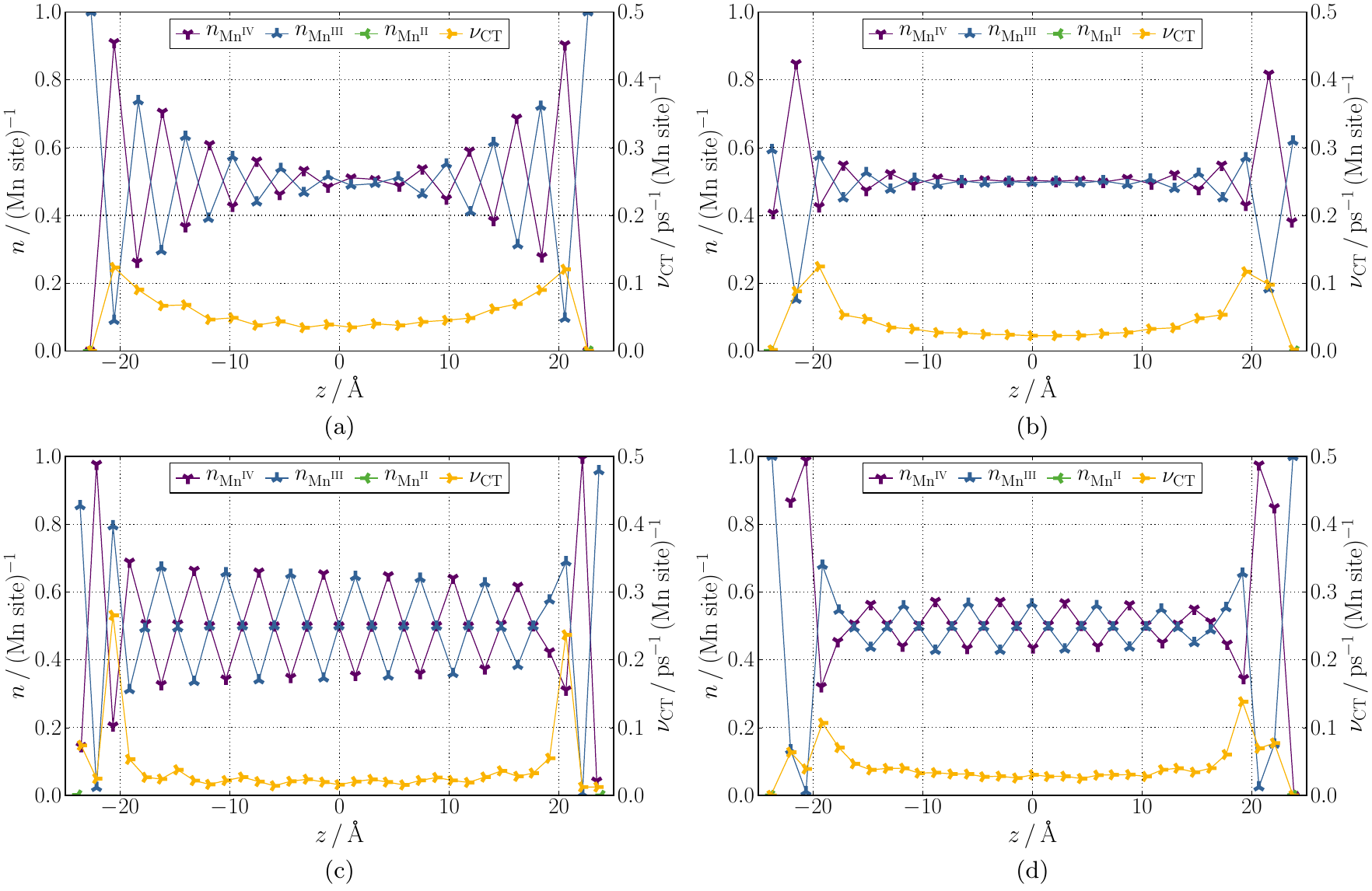}
\vspace*{-0.5cm}
\caption{Time averaged Mn oxidation state distributions and equilibrium charge transfer rates $\nu_\mathrm{CT}$ in the different layers of the (a) $\{100\}_\mathrm{Li}$, (b) $\{100\}_{\mathrm{Mn}_2\mathrm{O}_4}$, (c) $\{110\}_{\mathrm{LiMnO}_2}$, and (d) $\{110\}_{\mathrm{MnO}_2}$ Li$_x$Mn$_2$O$_4$-water interface systems. The oxidation state distribution is represented by the number of each species per Mn site $n$ in a layer. The lines are only shown to guide the eyes. The zero point of $z$ has been set to the center of the Li$_x$Mn$_2$O$_4$ slabs.}\label{fig:spin}
\end{figure*}

We start with the $\{100\}_\mathrm{Li}$ surface (Figure \ref{fig:structure} (a)), which is positively polarized due to Li termination, while the overall slab is neutral because of the charge compensation by the subsequent Mn$_2$O$_4$ layer containing formally a one-to-one ratio of Mn$^\mathrm{III}$ and Mn$^\mathrm{IV}$ ions. We note that due to the structurally identical surfaces at both sides of the slab, the system in total contains 23 Li layers and only 22 Mn$_2$O$_4$ layers resulting in a slight overall excess of Mn$^\mathrm{III}$ compared to Mn$^\mathrm{IV}$ ions in the system. Such a situation is not unphysical but typical for Li$_x$Mn$_2$O$_4$, e.g., for different loads of Li ions compensated by different oxidation states of Mn.

In our simulations we observe that in the outermost Mn$_2$O$_4$ layer on each side almost only Mn$^\mathrm{III}$ ions are present (Figure \ref{fig:spin} (a)). The reason for the preference of Mn$^\mathrm{III}$ ions in the topmost layer is the undercoordination of the Mn ions at interface sites by only five O$^{2-}$ ions compared to the octahedral coordination in the bulk material. We have not observed any long-living OH$^-$ ions formed by dissociation of water molecules, Consequently, significant protonation of interfacial O$^{2-}$ ions and adsorption of hydroxide ions at the undercoordinated Mn ions does not take place. Due to this lack of negative ions, the formation of Mn$^\mathrm{III}$ ions is favored at the interface and Mn$^\mathrm{IV}$ ions are predominantly found in the second Mn$_2$O$_4$ layer (Figure \ref{fig:spin} (a)). Deeper layers contain a decreasingly pronounced alternating excess of either Mn$^\mathrm{III}$ or Mn$^\mathrm{IV}$ ions, respectively, which becomes small after several layers. Mn$^\mathrm{II}$ ions are very rarely observed and, if found, they only emerge in the topmost layer with about $10^{-6}$ Mn$^\mathrm{II}$ ions per Mn site.

\begin{figure}[htb!]
\centering
\includegraphics[width=\columnwidth]{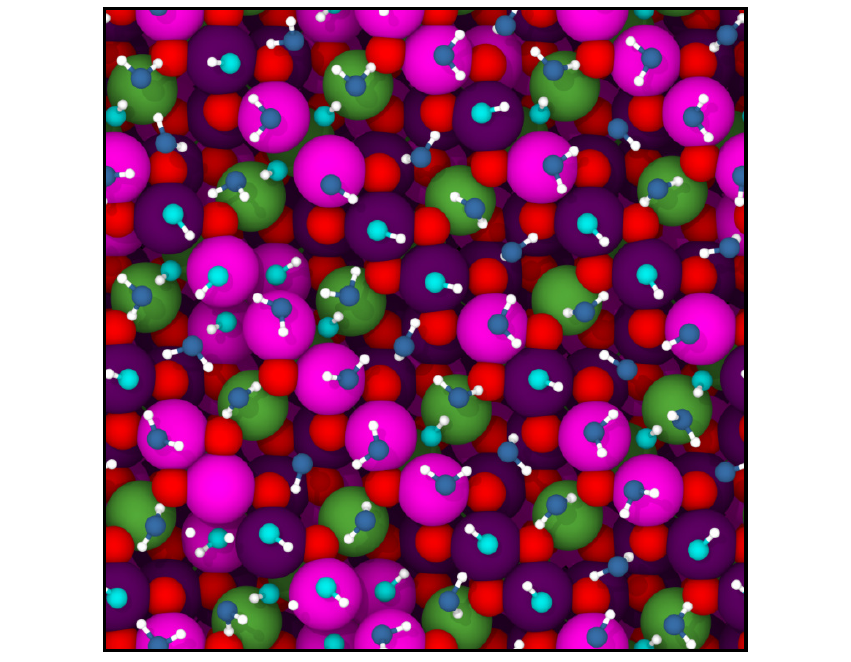}
\caption{Equilibrated example structure of the $\{100\}_{\mathrm{Mn}_2\mathrm{O}_4}$ Li$_x$Mn$_2$O$_4$-water interface including the solid surface and the first layer of the liquid projected on the $xy$ plane. The colors are according to the definition in Figure \ref{fig:structure} except that Mn$^\mathrm{III}$ ions are highlighted in pink, while Mn$^\mathrm{IV}$ ions are shown in violet.}\label{fig:structure_oxidation_states}
\end{figure}

In contrast to the $\{100\}_\mathrm{Li}$ system, both sides of the slab representing the $\{100\}_{\mathrm{Mn}_2\mathrm{O}_4}$ interface (Figure \ref{fig:structure} (b)) are terminated by Mn$_2$O$_4$ resulting in total in 22 Li layers and 23 Mn$_2$O$_4$ layers. The reduced amount of Li$^+$ ions in the $\{100\}_{\mathrm{Mn}_2\mathrm{O}_4}$ system compared to the $\{100\}_\mathrm{Li}$ system leads to a slight excess of Mn$^\mathrm{IV}$ ions instead of Mn$^\mathrm{III}$ ions. In the topmost layer the fraction is about three Mn$^\mathrm{III}$ to two Mn$^\mathrm{IV}$ ions (Figure \ref{fig:spin} (b)). The increased stability of Mn$^\mathrm{IV}$ ions in the topmost layer compared to the $\{100\}_\mathrm{Li}$ system can be explained by the coordination by OH$^-$ ions. For this system we find that OH$^-$ ions are adsorbed on top of about 43{\%} of the interface Mn ions. The majority of the corresponding protons formed in the dissociation of water molecules is attached to interface O$^{2-}$ ions, which are covered to about 21{\%} forming OH$^{-}$. This value is about half of the Mn coverage by OH$^-$ ions because there are twice as many interface O$^{2-}$ ions as Mn ions per layer. The OH$^-$ ions of the first liquid layer are preferably but not exclusively located at Mn$^\mathrm{IV}$ sites while Mn$^\mathrm{III}$ sites are typically coordinated by H$_2$O molecules (Figure \ref{fig:structure_oxidation_states}). The proposed intermediate state of the OER, in which two OH$^-$ ions are found on top of an [Mn$^\mathrm{III}_2$Mn$^\mathrm{IV}_2$O$_4$]$^{6+}$ unit,\cite{Cady2015} is therefore a rare configuration in our simulations explaining the low OER activity of the stoichiometric LiMn$_2$O$_4$ spinel.

The second layer on each side adapts to the Mn$^\mathrm{IV}$ excess in the system (Figure \ref{fig:spin} (b)). The alternating excess of either Mn$^\mathrm{III}$ or Mn$^\mathrm{IV}$ ions decays after a few layers converging to equal fractions of Mn$^\mathrm{III}$ and Mn$^\mathrm{IV}$ ions in deeper layers. In contrast to the $\{100\}_\mathrm{Li}$ system in which basically no Mn$^\mathrm{II}$ ions form, in the $\{100\}_{\mathrm{Mn}_2\mathrm{O}_4}$ system we find about $5\cdot10^{-4}$ Mn$^\mathrm{II}$ ions per Mn site in the topmost Mn$_2$O$_4$ layer with some statistical fluctuations depending on the surface and simulation. Substantially longer simulations are expected to be required to obtain fully converged values for such small fractions. As both sides of the Li$_x$Mn$_2$O$_4$ slab in every simulation are equal, converged results in Figure \ref{fig:spin} are symmetric with respect to $z=0$. Deviations from this behavior, which are small for most of our results, can consequently be employed to estimate the uncertainty caused by the finite simulation time.

The $\{110\}_{\mathrm{LiMnO}_2}$ (Figure \ref{fig:structure} (c)) and $\{110\}_{\mathrm{MnO}_2}$ (Figure \ref{fig:structure} (d)) interface systems show similar amounts of about $3\cdot10^{-4}$ and $4\cdot10^{-4}$ Mn$^\mathrm{II}$ ions per Mn sites in the topmost surface layers. While the $\{110\}_{\mathrm{LiMnO}_2}$ system contains a slight excess of Mn$^\mathrm{III}$ over Mn$^\mathrm{IV}$ ions, the opposite is the case for the $\{110\}_{\mathrm{MnO}_2}$ system. The topmost layer is still dominated by Mn$^\mathrm{III}$ ions in both systems with a ratio of nine Mn$^\mathrm{III}$ to one Mn$^\mathrm{IV}$ at the $\{110\}_{\mathrm{LiMnO}_2}$ interface (Figure \ref{fig:spin} (c)) and almost only Mn$^\mathrm{III}$ ions at the $\{110\}_{\mathrm{MnO}_2}$ interface (Figure \ref{fig:spin} (d)). The reason for this preference of Mn$^\mathrm{III}$ ions is the low coordination of the topmost Mn ions by only four O$^{2-}$ ions. As a consequence, the second layer in the $\{110\}_{\mathrm{LiMnO}_2}$ system and even the second and third layers in the $\{110\}_{\mathrm{MnO}_2}$ interface are predominantly occupied by Mn$^\mathrm{IV}$ ions. The influence of the interface on the oxidation state distribution vanishes already after about four layers. Regular oscillations of the alternating excess of either Mn$^\mathrm{III}$ or Mn$^\mathrm{IV}$, respectively, can be observed in the bulk (Figure \ref{fig:spin} (c) and (d)). As the amplitude of these oscillations varies for the $\{110\}_{\mathrm{LiMnO}_2}$ and $\{110\}_{\mathrm{MnO}_2}$ systems different arrangements of Mn$^\mathrm{III}$ and Mn$^\mathrm{IV}$ ions can lead to local minimum configurations. Further, due to electron hopping processes, the Mn$^\mathrm{III}$ and Mn$^\mathrm{IV}$ arrangements are dynamical at 298\,K and different local motifs are observed during the simulations even for the same interface systems. We note that the Mn oxidation state distribution is averaged over (001) planes in Figures \ref{fig:spin} (a) and (b) and over (110) planes in Figures \ref{fig:spin} (c) and (d) yielding different slices of the structure.

To compensate for the undercoordination of Mn ions by O$^{2-}$, about 94{\%} of the $\{110\}_{\mathrm{LiMnO}_2}$ interface Mn ions are in addition coordinated by approximately two OH$^-$ ions. These OH$^-$ ions are placed on bridge sites thus resembling the octahedral coordination in bulk Li$_x$Mn$_2$O$_4$ and are shared by two Mn ions (see Section \ref{sec:interface_structure}). About 46{\%} of the interface O$^{2-}$ ions contain adsorbed H$^+$ ions forming OH$^-$. For the $\{110\}_{\mathrm{MnO}_2}$ interface this value further increases slightly to about 48{\%}, while there is about one adsorbed OH$^-$ ion per interfacial Mn atom. In this case, the OH$^-$ ions originating from the liquid phase are not placed at Mn bridge sites but adopt empty octahedral coordination sites (see section \ref{sec:interface_structure}). To compare the OH$^-$ coverage between the $\{100\}_{\mathrm{Mn}_2\mathrm{O}_4}$ interface and the $\{110\}_{\mathrm{LiMnO}_2}$ and $\{110\}_{\mathrm{MnO}_2}$ interfaces, the number of Mn ions per interface area has to be taken into account. This number is about 1.5 times higher at the $\{100\}_{\mathrm{Mn}_2\mathrm{O}_4}$ interface compared to the other two interfaces. Still, the OH$^-$ coverage is higher at the $\{110\}_{\mathrm{LiMnO}_2}$ and $\{110\}_{\mathrm{MnO}_2}$ interfaces. Hence, the dissociation degree of H$_2$O molecules is higher at these two interfaces, which provide more empty octahedral coordination sites of the interfacial Mn ions than the $\{100\}_{\mathrm{Mn}_2\mathrm{O}_4}$ interface.

In addition to the Mn oxidation state distribution, we estimated the charge transfer rates related to electron hopping between the Mn$^\mathrm{II}$, Mn$^\mathrm{III}$, and Mn$^\mathrm{IV}$ ions. These rates were determined from the number of oxidation state changes per time. Because two oxidation states are changed by one electron hop, this number is divided by two. The trajectory data was collected every 0.1\,ps. The data in Figures \ref{fig:spin} (a) to (d) show that even the fastest processes are more than an order of magnitude slower than this sampling interval. This time scale difference ensures that the major fraction of processes is counted. To exclude counting of unsuccessful attempts to change the oxidation state, a transition is only considered in case the spin value of a Mn$^\mathrm{IV}$ ion increases above $1.9\,\hbar$, the spin value of a Mn$^\mathrm{III}$ ion increases above $2.3\,\hbar$ or decreases below $1.7\,\hbar$, or the spin value of a Mn$^\mathrm{II}$ ion decreases below $2.1\,\hbar$.

For all systems, the charge transfer rates are found to be largest close to the interface, with a maximum typically in the second to fourth Mn containing layer (Figures \ref{fig:spin} (a) to (d)). In the center of the solid slab the corresponding values are between about 0.02 to 0.03 charge transfers per ps and Mn site and hence much smaller than at the interface. These values are close to the value of 0.02 charge transfers per ps and Mn site obtained in our previous study of bulk LiMn$_2$O$_4$.\cite{Eckhoff2020b} The rates in the topmost layer of the $\{110\}_{\mathrm{LiMnO}_2}$ interface are higher than those in the topmost layers of the other interfaces, and in general lower charge transfer rates are found in layers with predominant single Mn oxidation states. For the $\{110\}_{\mathrm{LiMnO}_2}$ interface the rates are smaller in the second layer, which corresponds to the topmost layer at the $\{110\}_{\mathrm{MnO}_2}$ interface. Therefore, the charge transfer rates in the LiMnO$_2$ layers seem to be higher than those in the MnO$_2$ layers.

The electrons are not explicitly included in the HDNNP and the HDNNS identifies different oxidation states based on the local structural environment like, for example, the presence or absence of Jahn-Teller distortions. Consequently, the inhomogeneous distribution of Mn$^\mathrm{III}$ and Mn$^\mathrm{IV}$ ions in the systems as well as the electron hopping processes raise the question if the overall numbers of these ions are conserved during the simulations. In principle, in the solid phase there has to be a one-to-one ratio between the number of Li$^+$ ions and the number of Mn e$_\mathrm{g}$ electrons. Since the average number of Mn$^\mathrm{II}$ ions contributing two e$_\mathrm{g}$ electrons is rather small or even negligible at all interfaces, we expect to find about the same number of Li$^+$ ions and Mn$^\mathrm{III}$ ions, each containing one e$_\mathrm{g}$ electron, in the system. Indeed we observe that the number of e$_\mathrm{g}$ electrons stays approximately constant during all interface simulations (see Supplementary Material Figures S2 (a) to (d)). This observation provides evidence of the conservation of total charge and number of electrons and confirms the consistent description of the systems by the HDNNP. Only small fluctuations in the predictions are observed, since electron hopping processes can give rise to intermediate structures in which the geometry-based assignment of the oxidation state is unavoidably physically ambiguous.\cite{Eckhoff2020b} In addition, remaining prediction errors of the HDNNS may contribute to these fluctuations as well.

Due to the slight excess of Li$^+$ ions in the $\{100\}_\mathrm{Li}$ system related to the surface geometry, there are more Mn$^\mathrm{III}$ than Mn$^\mathrm{IV}$ ions present in this system. About 409 Mn$^\mathrm{III}$ and 383 Mn$^\mathrm{IV}$ ions are predicted by the HDNNS on average over the full simulation time of all three $\{100\}_\mathrm{Li}$ interface simulations, which contain all the same number of Mn ions. Since 414 Li$^+$ ions are present, the error in the number of e$_\mathrm{g}$ electrons obtained from the HDNNS prediction is only about 1.2{\%}. In contrast to the $\{100\}_\mathrm{Li}$ system, the $\{100\}_{\mathrm{Mn}_2\mathrm{O}_4}$ system contains more Mn$^\mathrm{IV}$ than Mn$^\mathrm{III}$ ions. Also in this case the HDNNS prediction is very accurate yielding about 400 Mn$^\mathrm{III}$ and 428 Mn$^\mathrm{IV}$ ions on average. Compared to the number of 396 Li$^+$ ions, the prediction error in the number of e$_\mathrm{g}$ electrons is again small (1.1{\%}). For the $\{110\}_{\mathrm{LiMnO}_2}$ system 401 e$_\mathrm{g}$ electrons are predicted compared to 408 Li$^+$ ions resulting in an underestimation of 1.8{\%}. Again, the excess of Mn$^\mathrm{III}$ over Mn$^\mathrm{IV}$ ions is predicted correctly (401:391). For the $\{110\}_{\mathrm{MnO}_2}$ system, for which 386 Mn$^\mathrm{III}$ and 406 Mn$^\mathrm{IV}$ ions are predicted by the HDNNS, an even better agreement with a deviation of only 0.4{\%} is reached, as this slab contains 384 Li$^+$ ions. In conclusion, the Mn oxidation states of all systems identified by the HDNNS are very accurately described via the geometric atomic environments of the Mn ions determined by the HDNNP energy surface.

The Mn$^\mathrm{II}$ ions in all systems have an above-average distance from the solid and are slightly displaced towards the liquid, which can be reasoned by the larger size of Mn$^\mathrm{II}$ ions. The Mn$^\mathrm{II}$ ions are preferably coordinated by H$_2$O instead of OH$^-$ of the water contact layer and the Mn-O distances are on average larger than for Mn$^\mathrm{III}$ and Mn$^\mathrm{IV}$ ions. Dissolution of Mn$^\mathrm{II}$ ions was not observed in the 5\,ns MD simulations at 298\,K and 1\,bar employing atomically flat solid surfaces without defects and in the absence of external electric fields. Therefore, dissolution seems to be rare in equilibrium under standard conditions. Especially, the inclusion of steps and defects at the solid surface is expected to increase the dissolution rate.\cite{Dove2005} Moreover, possible surface reconstructions as proposed for the \{110\} and \{111\} surfaces\cite{Hirayama2010, Benedek2011, Karim2013, Kim2015} and the formation of surface layers of different stoichiometry such as Mn$_3$O$_4$\cite{Tang2014, Amos2016, Gao2019} might also be relevant for Mn$^\mathrm{II}$ dissolution. In particular, the Mn$_3$O$_4$ tetragonal spinel structure, in which Mn$^\mathrm{II}$ ions substitute the Li$^+$ ions at the tetrahedral sites of Li$_x$Mn$_2$O$_4$, is an interesting candidate for the formation of dissolved Mn$^\mathrm{II}$ ions. These Mn$^\mathrm{II}$ ions in addition block the Li$^+$ channels in the spinel structure and thus need to be removed during charging of the battery.

In summary, the weak interaction between water and the $\{100\}_\mathrm{Li}$ interface seems to be responsible for only very little formation of Mn$^\mathrm{II}$ ions. The outermost Li$^+$ layer separates the water molecules from the Mn and O$^{2-}$ ions, which are important for the dissociation of water and the formation of long-living OH$^-$ ions. A high coordination by O$^{2-}$ ions, i.e., a more bulk-like environment, favors the formation of higher Mn oxidation states and leads to a weaker interaction with water. On the one hand, electron hopping and hence electrical conductivity is increased in the vicinity of the interface leading to higher battery performance when using smaller particles sizes. On the other hand, the formation of Mn$^\mathrm{II}$ ions is only observed close to the surface suggesting more durable battery materials when using larger particles. 

%%%%%%%%%%%%%%%%%%%%%%%%%%%%%%%%%%%%%%%%%%%%%%%%%%%%%%%%%%%%%%%%%%%%%%%%%%%%%%%%%%%%%%%%%%%%%%%%%%%%
\subsection{Structural characterization of the interfaces}\label{sec:interface_structure}
%%%%%%%%%%%%%%%%%%%%%%%%%%%%%%%%%%%%%%%%%%%%%%%%%%%%%%%%%%%%%%%%%%%%%%%%%%%%%%%%%%%%%%%%%%%%%%%%%%%%

The atomic structure as well as the reactivity of the interface are determined by the termination of the solid surface. For instance, we have seen that the Mn coordination can strongly affect the formation of OH$^-$ ions. The resulting degree of hydroxylation at the interface can be expected to be relevant for reactions at the surface such as the OER. Moreover, the structure and dynamics of the liquid in the vicinity of the interface and deviations of its properties from the bulk liquid are of high interest.

\begin{figure*}[htb!]
\centering
\includegraphics[width=\textwidth]{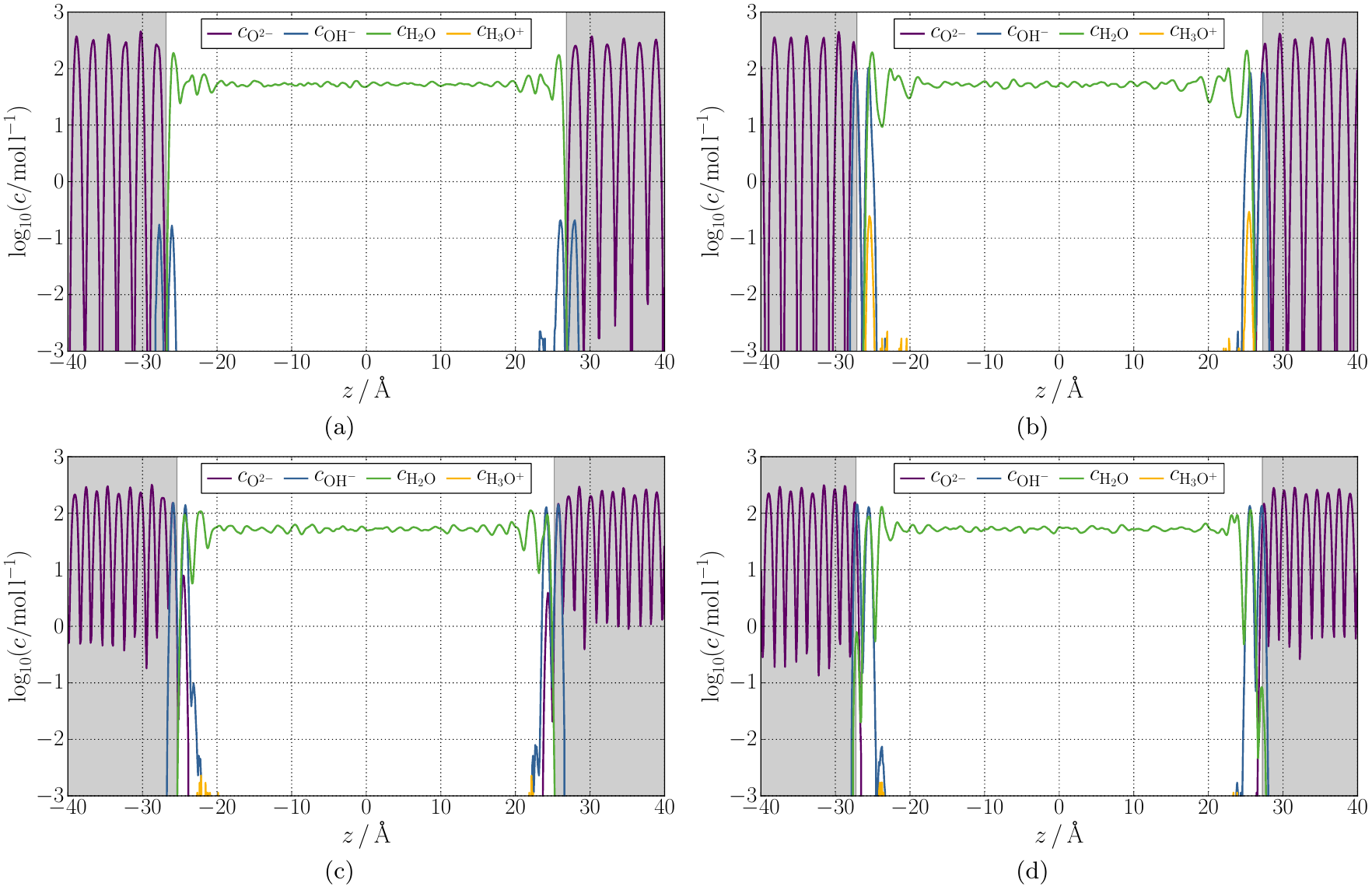}
\caption{Decadic logarithm of the time averaged concentration $c$ of different oxygen species as a function of the $z$ coordinate for the (a) $\{100\}_\mathrm{Li}$, (b) $\{100\}_{\mathrm{Mn}_2\mathrm{O}_4}$, (c) $\{110\}_{\mathrm{LiMnO}_2}$, and (d) $\{110\}_{\mathrm{MnO}_2}$ Li$_x$Mn$_2$O$_4$-water interface systems. The region of the Li$_x$Mn$_2$O$_4$ slab is highlighted by the gray background. The zero point of $z$ has been set to the center of the water slabs.}\label{fig:oxygen_species}
\end{figure*}

To assess the impact of the interface on the properties of the liquid, the system has to be sufficiently large to ensure the presence of a bulk-like region in the center of the liquid phase. This bulk-like region is not only important for comparing interfacial properties to those of the bulk, but also to obtain converged data for the interfacial properties. Figures \ref{fig:oxygen_species} (a) to (d) show the averaged atomic distributions in all four systems. The central region of the liquid phase is very similar for all different surfaces and shows only small fluctuations, which are much larger in the vicinity of the surfaces. The density of H$_2$O in the central 5\,{\AA} slice of the liquid phase is 0.946\,kg\,l$^{-1}$ ($c_{\mathrm{H}_2\mathrm{O}}=52.5\,\mathrm{mol\,l}^{-1}$). This density agrees very well with the value of 0.947\,kg\,l$^{-1}$ obtained in HDNNP-driven simulations of bulk water. Further, the properties of the hydrogen bond network are very similar in the centers of the liquid in all simulations (Supplementary Material). Consequently, the simulation cells, which all have a water region with a diameter of at least 50\,{\AA}, are large enough to yield bulk properties in the center, which is in excellent agreement with previous studies on other solid-water interfaces.\cite{Natarajan2016, Quaranta2017} An underestimation of the density compared to the experimental value of 0.997\,kg\,l$^{-1}$ at 298\,K and 1\,bar\cite{Wagner2002} is common in DFT calculations and was also observed in a previous study of water yielding 0.94\,kg\,l$^{-1}$\cite{Cheng2019} based on the revPBE0-D3\cite{Zhang1998, Adamo1999} DFT functional -- we note that our results are based on the PBE0r-D3\cite{Sotoudeh2017, Eckhoff2019} DFT functional.

The concentration profile as a function of the distance from the surfaces, which is proportional to the density profile with the molar mass as proportionality constant, shows two distinct OH$^-$ peaks of about the same size for each phase boundary (Figure \ref{fig:oxygen_species} (a) to (d)). These peaks correspond to protonated O$^{2-}$ ions of the solid and OH$^-$ ions adsorbed to Mn sites, respectively. Since these two peaks dominate the OH$^-$ concentration profile and are of about equal height, essentially all protons and hydroxide ions formed in the dissociation of water molecules are bound at the solid surface. The OH$^-$ concentration is about three orders of magnitude smaller at the $\{100\}_\mathrm{Li}$ interface compared to the other systems. The reason is that the OH$^-$ ions are not long-living at the $\{100\}_\mathrm{Li}$ interface. 

\begin{figure*}[htb!]
\centering
\includegraphics[width=\textwidth]{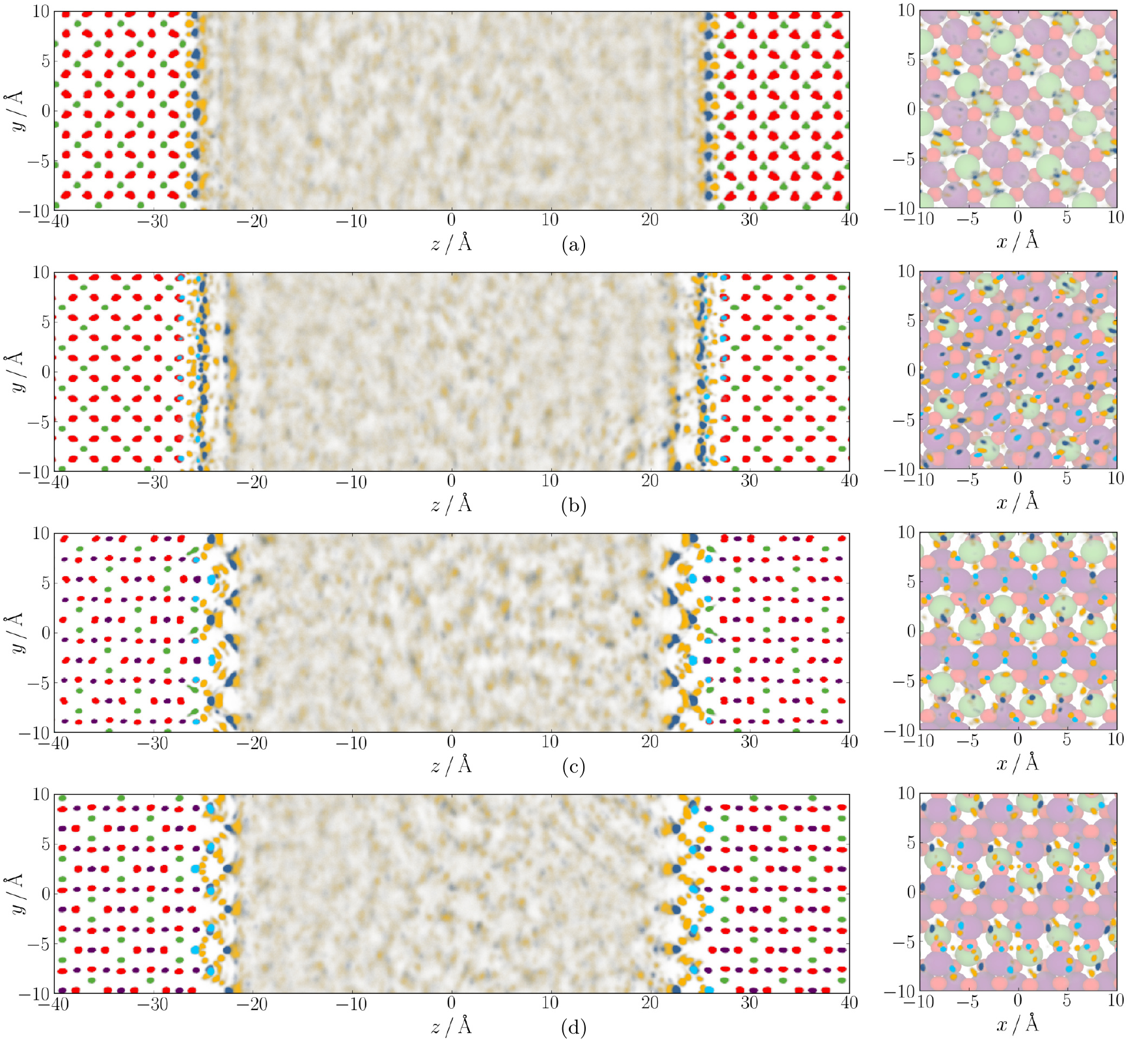}
\caption{Spatial atomic distributions projected onto the $zy$ plane of the (a) $\{100\}_\mathrm{Li}$, (b) $\{100\}_{\mathrm{Mn}_2\mathrm{O}_4}$, (c) $\{110\}_{\mathrm{LiMnO}_2}$, and (d) $\{110\}_{\mathrm{MnO}_2}$ Li$_x$Mn$_2$O$_4$-water interface systems are shown on the left. On the right the spatial atomic distributions up to 2.5\,{\AA} from the interface projected onto the $xy$ plane are represented with a ball model of the non-equilibrated vacuum solid surface in the background for reference. The time averaged concentration of each atomic species is represented by a linearly increasing opacity. In the case of oxygen, each of the species differing in hydrogen content is plotted individually. The spatial distributions are stacked on top of each other in the order from bottom to top H (yellow), Li (green), Mn (violet), O$^{2-}$ (red), O in H$_2$O (blue), and O in OH$^-$ (turquoise). The spatial distributions of Li, Mn, and O$^{2-}$ are not shown in the $xy$ projections.}\label{fig:layer_distribution}
\end{figure*}

Oscillations of the H$_2$O concentration in the vicinity of the $\{100\}_{\mathrm{Mn}_2\mathrm{O}_4}$ interface are more pronounced than in the vicinity of the $\{100\}_\mathrm{Li}$ interface (Figure \ref{fig:oxygen_species} (a) and (b)). The depletion layer beyond the first water peak can be observed in the spatial atomic distributions in Figure \ref{fig:layer_distribution} (b) as well. The relatively high concentration of H$_3$O$^+$ ions at the same distance as the second OH$^-$ peak will be discussed in Section \ref{sec:proton_transfer}.

\begin{figure}[htb!]
\centering
\includegraphics[width=\columnwidth]{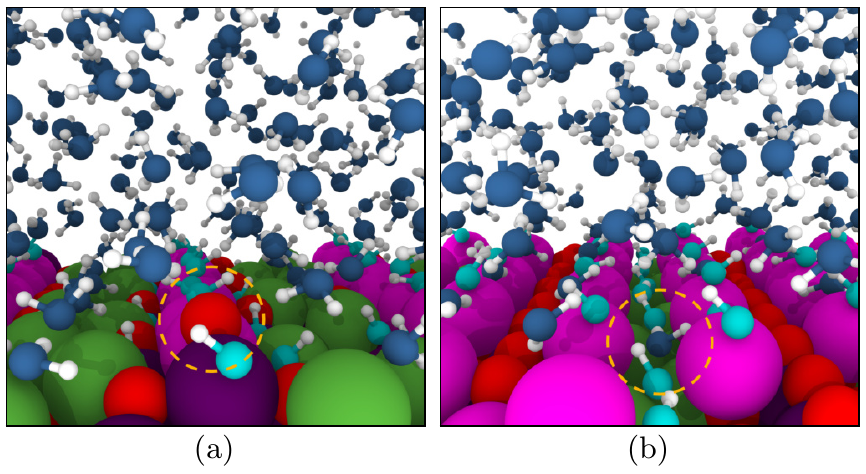}
\caption{Equilibrated example structures of (a) an O$^{2-}$ ion (red) formed in an additional layer at the $\{110\}_{\mathrm{LiMnO}_2}$ interface and (b) an H$_2$O molecule (blue) formed in the topmost solid layer of the $\{110\}_{\mathrm{MnO}_2}$ interface. Both species are highlighted by a yellow circle.}\label{fig:structure_interface}
\end{figure}

The structural deviation of the contact layer from the bulk liquid for the $\{110\}_{\mathrm{LiMnO}_2}$ and $\{110\}_{\mathrm{MnO}_2}$ systems is even more pronounced. In the case of the $\{110\}_{\mathrm{LiMnO}_2}$ system, the OH$^-$ concentration profile shows the presence of OH$^{-}$ ions rather far from the surface. The formation of a small additional O$^{2-}$ peak on top of the surface with maximum concentrations between 1 and 10\,mol\,l$^{-1}$ is particularly interesting. This peak implies that some H$_2$O molecules can be deprotonated twice to partially complete the octahedral coordination of the Mn ions as shown in Figure \ref{fig:structure_interface} (a). This formation of surface exposed O$^{2-}$ ions is potentially of interest for catalytic reactions due to their low coordination. Furthermore, a Mn$^\mathrm{IV}$ ion is often found close to a surface exposed O$^{2-}$ ion although the first solid layer is typically dominated by Mn$^\mathrm{III}$ ions at this surface. Thus, the surface exposed O$^{2-}$ ions have an impact on the distribution of Mn$^\mathrm{III}$ and Mn$^\mathrm{IV}$ ions.

At the $\{110\}_{\mathrm{MnO}_2}$ interface an H$_2$O concentration of 0.1 to 1\,mol\,l$^{-1}$ is found in the topmost layer of the solid surface, which implies the opposite process. Here, O$^{2-}$ ions are protonated twice thus forming water as shown in Figure \ref{fig:structure_interface} (b). A correlation of Mn$^\mathrm{II}$ formation to the formation of H$_2$O molecules in the topmost solid layer is not observed. Mn$^\mathrm{II}$ ions are most often observed in environments in which the Mn ion is coordinated by two H$_2$O molecules from above instead of typically one OH$^-$ ion and one H$_2$O molecule. The double deprotonation and protonation processes can be viewed as surface reconstructions of the $\{110\}$ Li$_x$Mn$_2$O$_4$ surfaces.

On the left side of Figures \ref{fig:layer_distribution} (a) to (d) the time averaged spatial atomic distributions projected onto the $zy$ plane are shown for each interface. As expected the bulk solid has a regular pattern reflecting the crystal structure while the bulk liquid has a diffuse distribution in all simulations. However, at the different solid-liquid interfaces the liquid phase shows various interesting strongly bound water structural features, which are less mobile due to the strong interaction with the surface. An adaption of the interfacial water layers to optimize the interaction to the solid as well as the hydrogen bond network to the bulk liquid has been observed for different metal surfaces as well and can yield very specific water structures depending on the solid surface.\cite{Liriano2017, Gerrard2019} The thickness of the strongly bound water layer depends on the underlying solid surface. For the $\{100\}_\mathrm{Li}$ interface the strongly bound water layer has a diameter of about 1.5 to 2\,{\AA} only, while it is 3 to 4\,{\AA} thick for the $\{100\}_{\mathrm{Mn}_2\mathrm{O}_4}$ and $\{110\}_{\mathrm{LiMnO}_2}$ interfaces. However, the structure at the $\{100\}_{\mathrm{Mn}_2\mathrm{O}_4}$ interface is dominated by a two-dimensional dense water layer on top of the solid, while the structure at the $\{110\}_{\mathrm{LiMnO}_2}$ interface is three-dimensional. A pattern similar to the latter one is also observed at the $\{110\}_{\mathrm{MnO}_2}$ interface. Here the strongly bound water layer is even 3.5 to 4.5\,{\AA}.

\begin{figure}[htb!]
\centering
\includegraphics[width=\columnwidth]{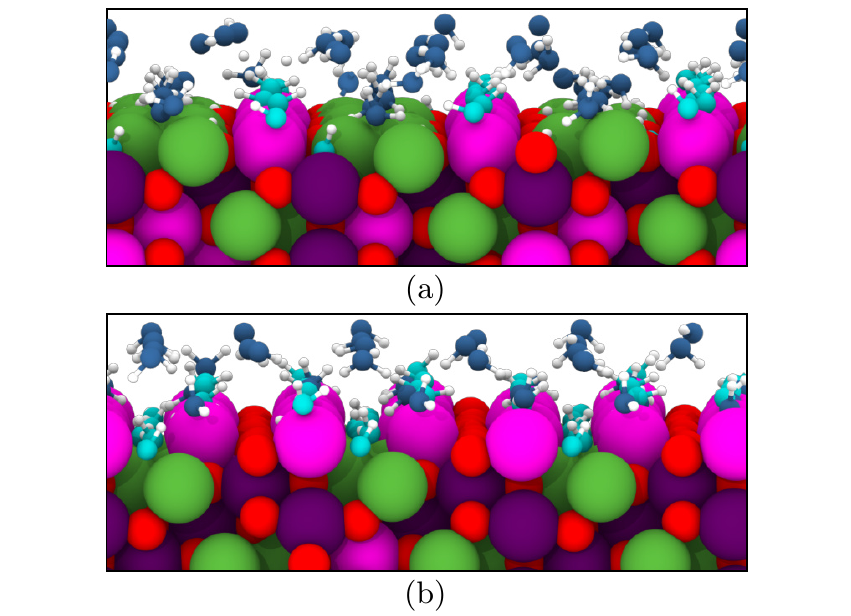}
\caption{Equilibrated example structures of the strongly bound water layers in the (a) $\{110\}_{\mathrm{LiMnO}_2}$ and (b) $\{110\}_{\mathrm{MnO}_2}$ Li$_x$Mn$_2$O$_4$-water interface systems.}\label{fig:structure_crystalline_water}
\end{figure}

To investigate the strongly bound water layers in more detail the right panels of Figures \ref{fig:layer_distribution} (a) to (d) show the spatial atomic distributions in water films of 2.5\,{\AA} diameter above the surface starting from the H atom closest to the surface projected onto the $xy$ plane. The small strongly bound water layer on top of the $\{100\}_\mathrm{Li}$ interface forms due to the attractive interactions between the oxygen of H$_2$O and Li$^+$ ions as well as due to hydrogen bond formation between the hydrogen of H$_2$O and O$^{2-}$ ions at the surface (Figure \ref{fig:layer_distribution} (a)). A pattern can be observed at the interface but the concentrations at the strongly bound water sites are lower than those for the other interfaces (lower opacity of the first liquid layer in the right panels of Figure \ref{fig:layer_distribution} (a) than in (b) to (d)). Consequently, the H$_2$O molecules at the $\{100\}_\mathrm{Li}$ interface are more mobile than the various water species at the other interfaces, corresponding to the weaker interaction between water and the $\{100\}_\mathrm{Li}$ surface discussed above.

As mentioned in Section \ref{sec:Mn_oxidation_state} OH$^-$ ions are formed at the $\{100\}_{\mathrm{Mn}_2\mathrm{O}_4}$, $\{110\}_{\mathrm{LiMnO}_2}$, and $\{110\}_{\mathrm{MnO}_2}$ interfaces and cover the solid surface to a large fraction (Figures \ref{fig:layer_distribution} (b) to (d)). These OH$^-$ ions are rather strongly bound to specific sites and thus less mobile than the water molecules. Moreover, they are able to form strong hydrogen bonds and can order the water molecules. Therefore, they have a large impact on the formation of the strongly bound water layer. The H$_2$O/OH$^-$ distribution in the first layer of the liquid phase does not follow a regular pattern at the $\{100\}_{\mathrm{Mn}_2\mathrm{O}_4}$ interface (Figure \ref{fig:layer_distribution} (b) as well as Figure \ref{fig:structure_oxidation_states}). Hence, the Mn$^\mathrm{III}$/Mn$^\mathrm{IV}$ distribution in the first layer of the solid phase is also disordered (Figure \ref{fig:structure_oxidation_states}).

The order in the H$_2$O/OH$^-$ distribution increases for the $\{110\}_{\mathrm{LiMnO}_2}$ and $\{110\}_{\mathrm{MnO}_2}$ interfaces. At the $\{110\}_{\mathrm{LiMnO}_2}$ interface most OH$^-$ ions of the first layer of the liquid bridge the Mn sites, which are arranged in rows in the first solid layer (e.g., $y\approx5\,\text{\AA}$ in Figure \ref{fig:layer_distribution} (c) as well as Figures \ref{fig:structure_interface} (a) and \ref{fig:structure_crystalline_water} (a)). Most H$_2$O molecules in the first layer are aligned in rows close to the Li$^{+}$ ions with H atoms pointing to O$^{2-}$ ions of the solid (e.g., $y\approx1\,\text{\AA}$ in Figure \ref{fig:layer_distribution} (c)). In the first layer of the solid rows of alternating O$^{2-}$ and OH$^-$ ions are formed (e.g., $y\approx-1\,\text{\AA}$ in Figure \ref{fig:layer_distribution} (c)). O$^{2-}$ excess can lead to Mn$^\mathrm{IV}$ ions in the topmost solid layer.

In the first solid layer of the $\{110\}_{\mathrm{MnO}_2}$ interface an alternating pattern of O$^{2-}$ and OH$^-$ rows is formed, whereby O$^{2-}$ ions bridge the underlying Mn sites and the OH$^-$ ions are above the Li sites (e.g., $y\approx3\,\text{\AA}$ and $y\approx7\,\text{\AA}$ in Figure \ref{fig:layer_distribution} (d) as well as Figures \ref{fig:structure_interface} (b) and \ref{fig:structure_crystalline_water} (b)). In some simulations the H$_2$O molecules and OH$^-$ ions in the first liquid layer are arranged in alternating rows (e.g., $x\approx2\mathrm{\ to\ }3\,\text{\AA}$ and $x\approx5\mathrm{\ to\ }6\,\text{\AA}$ in Figure \ref{fig:layer_distribution} (d)).

The oxygen atoms of the water species in the first liquid layer generally tend to continue the oxygen face centered cubic (fcc) lattice of the solid yielding an energetically favored coordination of the Li and Mn ions (see Supplementary Material Figures S7 (a) and (b) for representations of the oxygen lattice only). The second liquid layer is most pronounced in the case of the $\{110\}_{\mathrm{MnO}_2}$ interface. The view on the $yz$ plane still seems to agree with the fcc lattice (Figure \ref{fig:structure_crystalline_water} (b)) but a view on the $xz$ plane shows that the second layer does not match the fcc lattice (Supplementary Material Figure S8). The same observation is obtained for the $\{110\}_{\mathrm{LiMnO}_2}$ interface. However, the strongly bound water layer cannot be assigned to a specific lattice structure or ice polymorph, since in most cases the OH$^{-}$ ions at the surface do not form a sufficiently regular pattern.

In conclusion, with increasing OH$^-$ concentration complex structural patterns can form which also affect the oxidation states of the underlying Mn ions. Interactions between the solid and the liquid lead to favored positions of the atoms. The oxygen atoms of OH$^-$ ions and H$_2$O molecules in the contact layer of the liquid prefer to be at sites which complete a bulk-like octahedral coordination of Mn or tetrahedral coordination of Li. The respective OH$^-$ ions thereby favor to bind to interfacial Mn ions. Therefore, strongly bound water structures form at the interface.

To compare the residence lifetimes of the O atoms in the strongly bound water layers at the different interfaces as well as in the bulk quantitatively, the correlation function,
\begin{align}
C_\mathrm{res}(t)=\dfrac{1}{N_\mathrm{tot}-N(t)}\sum_{t_0=0}^{t_\mathrm{tot}-t}\dfrac{\mathbf{r}_\mathrm{init}(t_0)\cdot\mathbf{r}(t_0+t)}{|\mathbf{r}_\mathrm{init}(t_0)|}\ ,
\end{align}
can be used with the total number of simulation steps $N_\mathrm{tot}$ and total simulation time $t_\mathrm{tot}$. $N(t)$ is the number of steps until time $t$. The quantity,
\begin{align}
r_i(t)=\begin{cases}1&\mathrm{for}\ z_\mathrm{min}\leq z_i(t)\leq z_\mathrm{max}\\0&\mathrm{otherwise}\end{cases}\ ,
\end{align}
keeps track if the oxygen $z$ coordinate of atom $i$ at time $t$ is outside the interval $[z_\mathrm{min},z_\mathrm{max}]$. O atoms which are close to the boundary at the time $t_0$ can lead to a large impact of short-term boundary crossing. To reduce this effect, the vector $\mathbf{r}_\mathrm{init}$ can employ a slightly smaller interval $[z_\mathrm{min},z_\mathrm{max}]$ than used for $\mathbf{r}$.

The residence correlation function is 1 if all initially resident O atoms are in the interval and becomes 0 if none of these O atoms is left in the interval. This function enables to measure the residence lifetime of the O atoms in given slices of the simulation cell. The residence lifetime can be extracted by a biexponential fit using
\begin{align}
C_\mathrm{res}^\mathrm{fit}(t)=C_1\exp\left(-\dfrac{t}{\tau_1}\right)+(1-C_1)\exp\left(-\dfrac{t}{\tau_2}\right)\ .
\end{align}
$\tau_1$ is the residence lifetime due to long-range reorganization caused by diffusion and $\tau_2$ accounts for short-term crossing of the boundaries. The weights of both processes $C_1$ and $1-C_1$ add up to one. For verification we show in the Supplementary Material Figure S9 monoexponential fits excluding the initial drop of the correlation functions which yield similar residence lifetimes $\tau_1$.

\begin{figure}[htb!]
\centering
\includegraphics[width=\columnwidth]{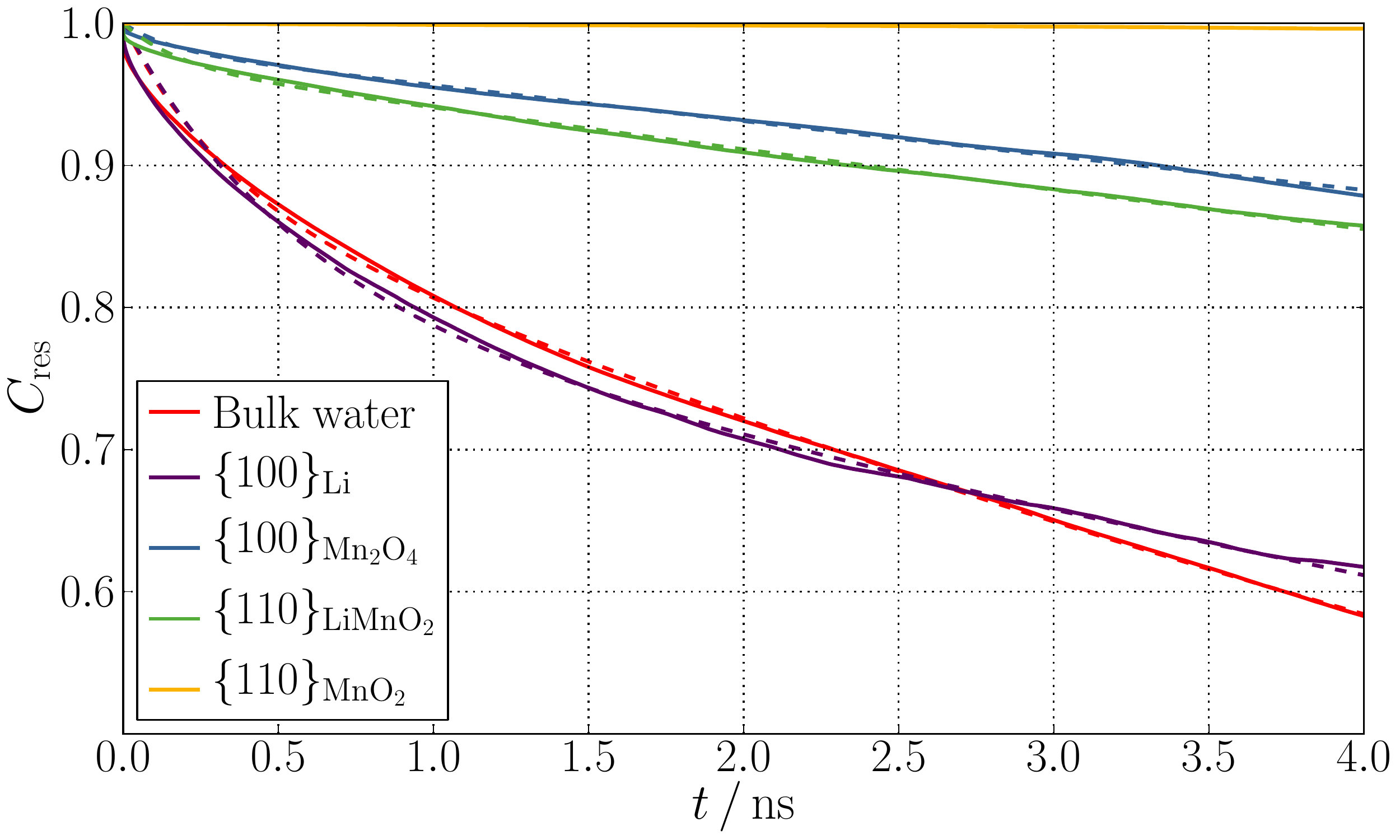}
\caption{Correlation functions $C_\mathrm{res}$ of the oxygen residence in the center of the liquid phase and in the interfacial liquid phase close to the solid surfaces. For the bulk case O atoms are considered which are located in the central 5\,{\AA} slice of the liquid phase at $t_0=0$. For the interface cases, O atoms in the two 2.5\,{\AA} slices at both surfaces of the solid at $t_0=0$ contribute. The contributions to the correlation function decrease from 1 to 0 if the O atoms leave the slices by more than 0.3\,{\AA} and they increase from 0 to 1 if the O atoms enter this region again. The results of all twelve simulations are averaged for the bulk water data. For the interface data the results of the three simulations for each interface are averaged. Dashed lines correspond to biexponential fits.}\label{fig:correlation_residence}
\end{figure}

The residence correlation functions of the central bulk water region and the different interface regions show that the long-range diffusion is faster in the bulk than at the interfaces (Figure \ref{fig:correlation_residence}). The residence lifetime of the O atoms in the central region is about 9\,ns. Since the $\{100\}_\mathrm{Li}$ interface has only a very thin strongly bound layer of H$_2$O molecules, the residence lifetime of about 14\,ns is still similar to bulk water, which is expected due to the weak interaction of this surface with the liquid. However, the $\{100\}_{\mathrm{Mn}_2\mathrm{O}_4}$ and $\{110\}_{\mathrm{LiMnO}_2}$ interfaces containing large amounts of OH$^-$ ions bind the hydroxide/strongly bound water layer leading to higher residence lifetimes of about 37 and 31\,ns, respectively. For the $\{110\}_{\mathrm{MnO}_2}$ interface only very few oxygen species in the strongly bound layer can escape from the interface during the 5\,ns simulation. We note that the relative comparison of these numbers is more valuable than the absolute numbers because the underlying DFT method may generally tend to under- or overstructure water leading to faster or slower diffusion than obtained experimentally.\cite{Lin2012, Gillan2016, Chen2017}

%%%%%%%%%%%%%%%%%%%%%%%%%%%%%%%%%%%%%%%%%%%%%%%%%%%%%%%%%%%%%%%%%%%%%%%%%%%%%%%%%%%%%%%%%%%%%%%%%%%%
\subsection{Proton transfer}\label{sec:proton_transfer}
%%%%%%%%%%%%%%%%%%%%%%%%%%%%%%%%%%%%%%%%%%%%%%%%%%%%%%%%%%%%%%%%%%%%%%%%%%%%%%%%%%%%%%%%%%%%%%%%%%%%

An atomistic understanding of site-specific reaction mechanisms is crucial in the bottom-up development of advanced heterogeneous catalysts. The reactivity of solid surfaces can be highly dependent on the interface structure. The identification of active solid surface sites can therefore target the design of particle shapes and sizes yielding a higher activity. For instance, we have seen that the dissociation of water molecules is much more frequent at the $\{100\}_{\mathrm{Mn}_2\mathrm{O}_4}$, $\{110\}_{\mathrm{LiMnO}_2}$, and $\{110\}_{\mathrm{MnO}_2}$ interfaces compared to the $\{100\}_\mathrm{Li}$ interface. Since the initial steps of the proposed OER mechanism at Li$_x$Mn$_2$O$_4$ include the formation of two OH$^-$ ions on top of [Mn$^\mathrm{III}_2$Mn$^\mathrm{IV}_2$O$_4$]$^{6+}$,\cite{Cady2015} PT reactions at the interface are likely to play an important role. 

\begin{figure*}[htb!]
\centering
\includegraphics[width=\textwidth]{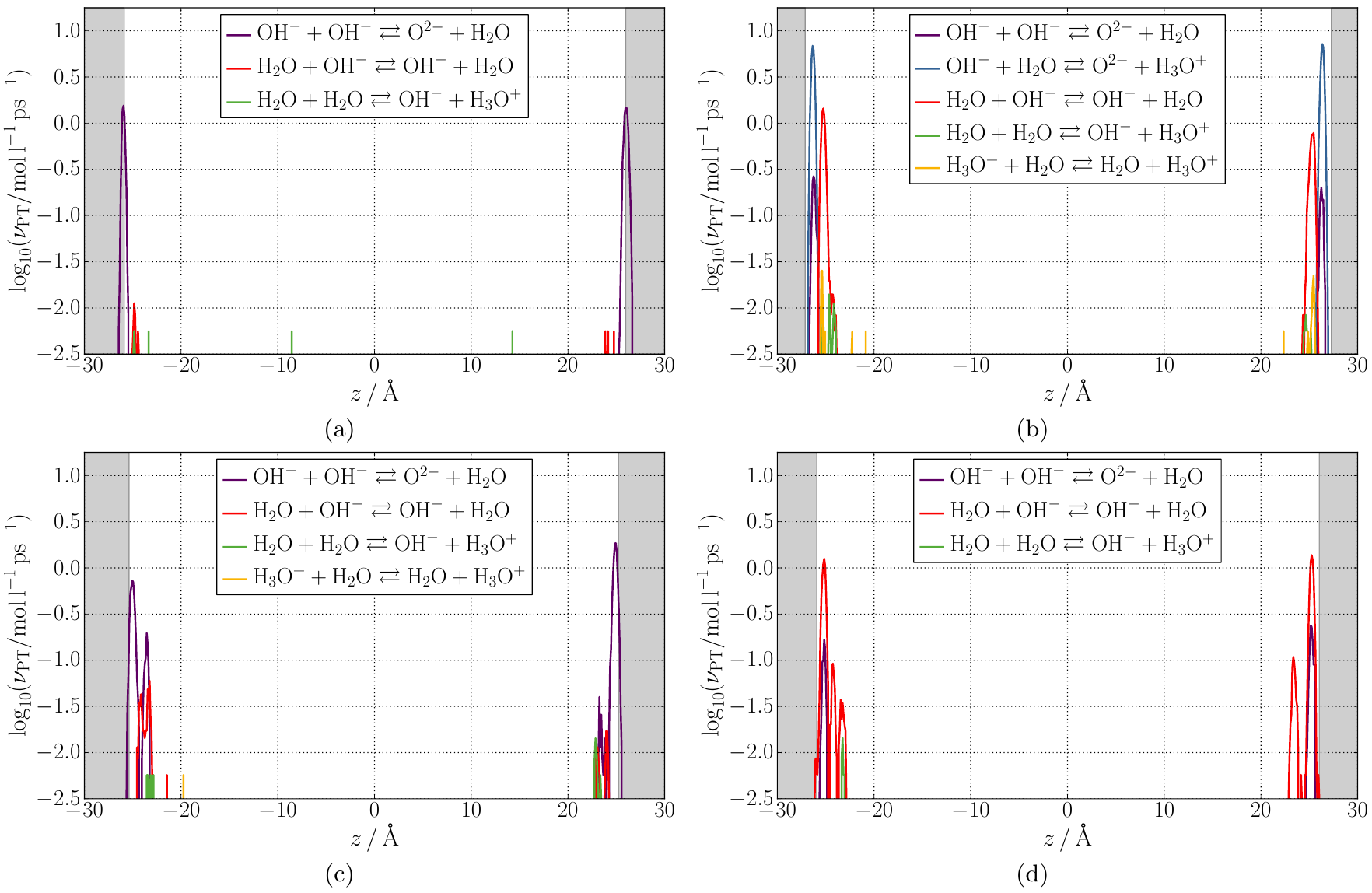}
\caption{Decadic logarithm of the PT reaction rates $\nu_\mathrm{PT}$ between different water species as a function of the mean $z$ coordinate of the transferred H atom for the (a) $\{100\}_\mathrm{Li}$, (b) $\{100\}_{\mathrm{Mn}_2\mathrm{O}_4}$, (c) $\{110\}_{\mathrm{LiMnO}_2}$, and (d) $\{110\}_{\mathrm{MnO}_2}$ Li$_x$Mn$_2$O$_4$-water interface systems. The solid phase is highlighted by a gray background.}\label{fig:PT_species}
\end{figure*}

\begin{figure}[htb!]
\centering
\includegraphics[width=\columnwidth]{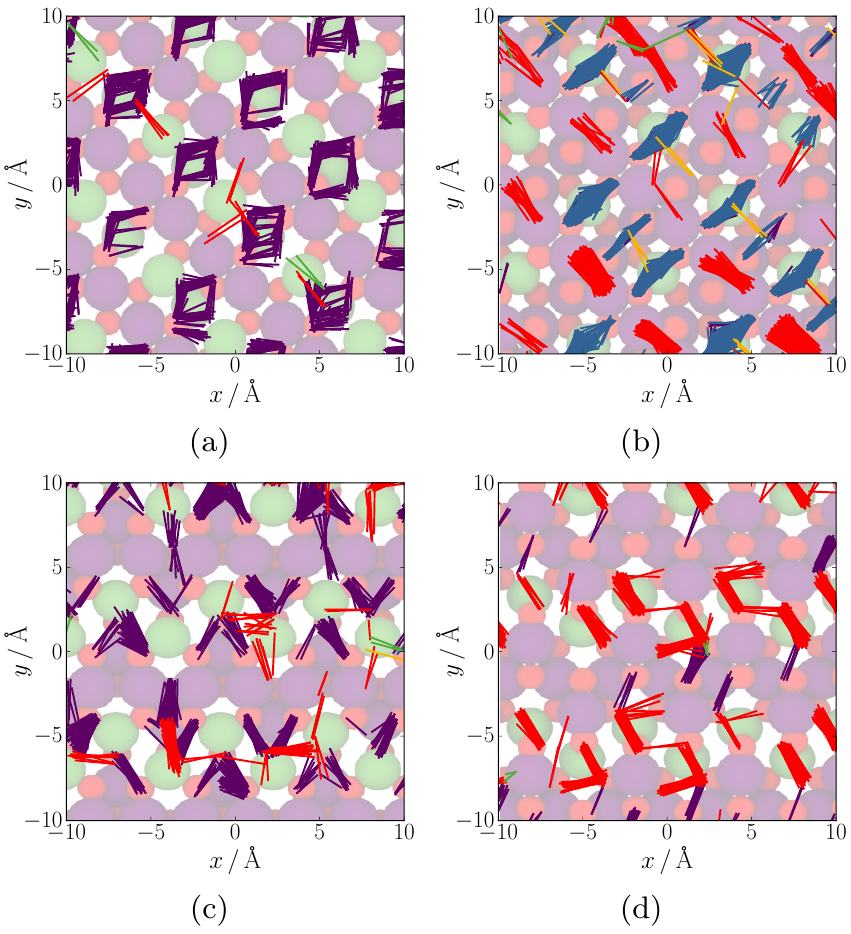}
\caption{Vectors connecting the oxygen atoms participating in PT events up to 7.5\,{\AA} from the interface projected onto the $xy$ plane of the (a) $\{100\}_\mathrm{Li}$, (b) $\{100\}_{\mathrm{Mn}_2\mathrm{O}_4}$, (c) $\{110\}_{\mathrm{LiMnO}_2}$, and (d) $\{110\}_{\mathrm{MnO}_2}$ Li$_x$Mn$_2$O$_4$-water interface systems. Missing translational symmetry is caused by limited sampling in the finite simulation time. A ball model of the ideal vacuum solid surface is shown in the background for reference. The colors represent the different water species participating in the PT and are defined in Figure \ref{fig:PT_species}.}\label{fig:PT_directions}
\end{figure}

\begin{figure}[htb!]
\centering
\includegraphics[width=\columnwidth]{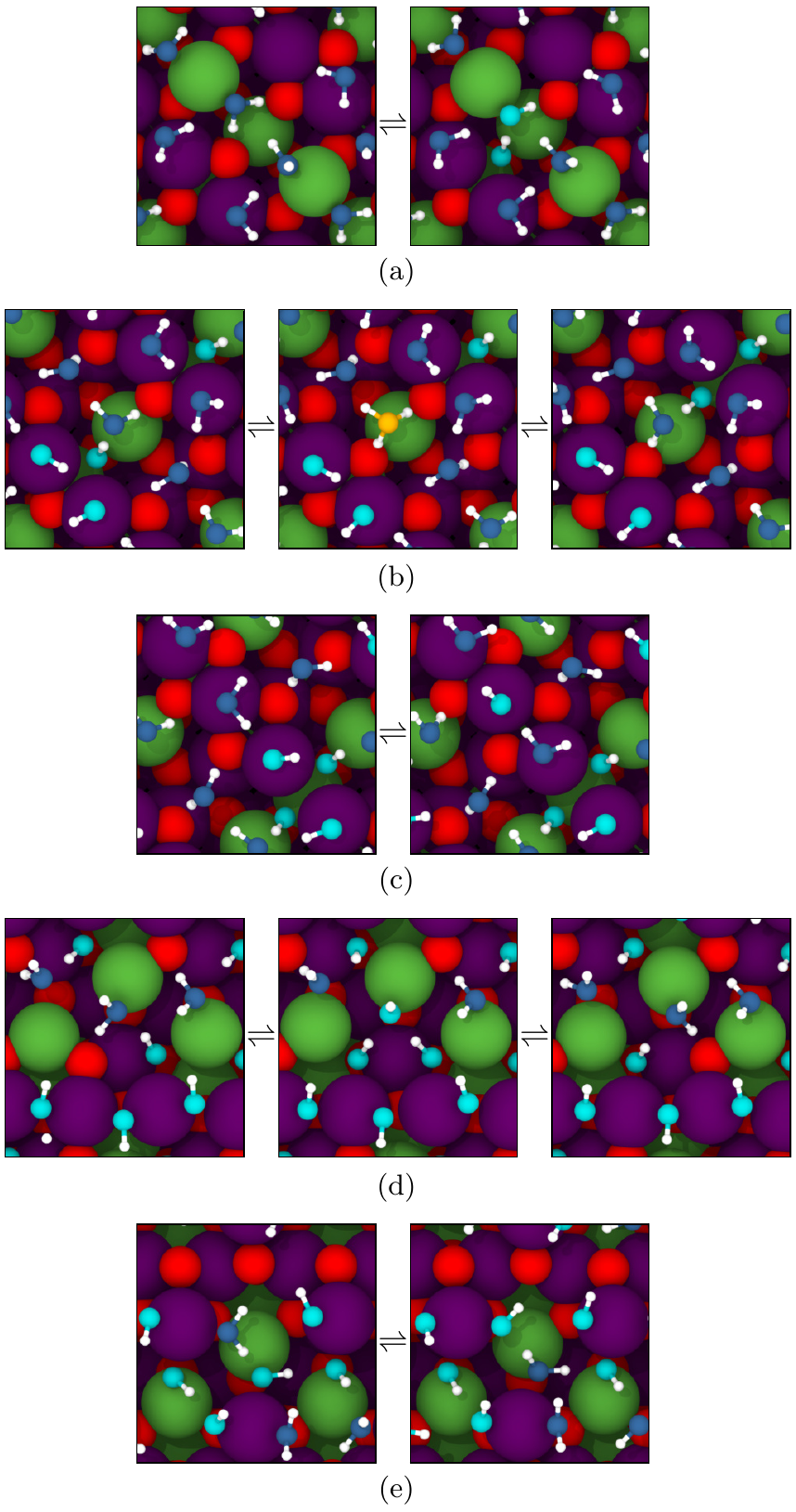}
\caption{Top views of the dominant PT reactions in the (a) $\{100\}_\mathrm{Li}$, (b) and (c) $\{100\}_{\mathrm{Mn}_2\mathrm{O}_4}$, (d) $\{110\}_{\mathrm{LiMnO}_2}$, and (e) $\{110\}_{\mathrm{MnO}_2}$ Li$_x$Mn$_2$O$_4$-water interface systems projected onto the $xy$ plane. In this figure all Mn oxidation states are colored violet.}\label{fig:PT_reactions}
\end{figure}

The PT reaction rates in Figures \ref{fig:PT_species} (a) to (d) are obtained from changes in the H atom assignments to the closest O atom, i.e., it is the number of assignment changes per unit time. The projected networks of the PT reactions near the interfaces are shown in Figures \ref{fig:PT_directions} (a) to (d). In the ideal case, all PT processes from one local minimum configuration to another one are counted, while proton rattling events, in which H atoms quickly jump back and forth between O atoms without settling on one of them, are excluded. However, an unambiguous distinction of both types is not possible. We decided to use a sampling interval of 0.1\,ps to reduce counting of proton rattling events because in this time interval a new local minimum can often be adopted. This approach yields a lower boundary of the absolute reaction rates since some PTs can be missed if forward and backward reactions happen rapidly. Also, nuclear quantum effects, which are not included in our simulations, are expected to increase the PT reaction rates.\cite{Litman2018} In the Supplementary Material we calculate the PT reaction rates of shorter trajectories using a reduced sampling interval of 1\,fs, while requiring that the assignment of the H atom to the closest O atom has to be maintained for 50\,fs after a transition to be counted as PT reaction. This approach yields mostly similar reaction rates of the frequent processes, while rare processes are sampled worse due to the shorter total simulation time (Supplementary Material Figures S10 (a) to (d)). Since the same approximations have been applied in the analysis of all PT processes, errors are expected to cancel to some extent resulting in comparable relative reaction rates. The PT reactions are classified by the water species involved in the reaction. In case a certain ion or molecule is participating in multiple PTs during a single sampling interval, the overall process is decomposed into the respective elementary steps. Since the PT rates of forward and backward reactions are equal in equilibrium according to the principle of detailed balance, their average has been calculated.

At the $\{100\}_\mathrm{Li}$ interface the dominant PT mechanism is the hydroxylation of the solid surface (Figure \ref{fig:PT_species} (a)), which has been called ``surface PT'' in earlier work.\cite{Quaranta2017, Hellstroem2019, Quaranta2019} An O$^{2-}$ ion from the solid accepts one proton from an H$_2$O molecule such that two OH$^-$ ions are formed (Figure \ref{fig:PT_reactions} (a)). The backward reactions most often happen between the same atoms (Figure \ref{fig:PT_directions} (a)) since the OH$^-$ ions are not long-living at the $\{100\}_\mathrm{Li}$ interface. PTs between H$_2$O molecules and OH$^-$ ions in the first liquid layer, i.e., ``adlayer PT'',\cite{Quaranta2017} happen rarely. Due to the effectively zero-dimensional proton transport by these PT reactions, the long-range transport is only given by diffusion and not by PT reactions following a Grotthus-like mechanism. In the liquid phase we can observe rare events of the autoionization of water, i.e., the formation of OH$^-$ and H$_3$O$^+$ ions from two H$_2$O molecules (very small green peaks between $-15\,\text{\AA}<z<15\,\text{\AA}$ in Figure \ref{fig:PT_species} (a)). These PT processes in the bulk can also be observed in some of the other simulations which are not shown in Figure \ref{fig:PT_species}. The OH$^-$ and H$_3$O$^+$ ions always recombine very quickly. This behavior is expected as a pH value of 7 in bulk water would mean that only a concentration of $10^{-7}\,\mathrm{mol\,l}^{-1}$ H$_3$O$^+$ ions are present. Consequently, there would be around 0.1 H$_3$O$^+$ ions in one of the $5\cdot10^4$ structures of a simulation trajectory including about $10^3$ H$_2$O molecules each.

The PT paths are completely different at the $\{100\}_{\mathrm{Mn}_2\mathrm{O}_4}$ interface. Here, the dominant mechanism is the transfer of one proton from a surface OH$^-$ ion to an H$_2$O molecule (Figure \ref{fig:PT_species} (b)). The H$_3$O$^+$ ion can be seen as transition state of the PT which rapidly transfers another proton to an O$^{2-}$ ion of the surface (Figure \ref{fig:PT_reactions} (b)). Such solvent-assisted PTs have also been observed at ZnO-water surfaces in previous work.\cite{Hellstroem2019} Surface PTs (violet) can be observed at the same locations as the solvent-assisted PTs (blue) in Figure \ref{fig:PT_directions} (b). Hence, the hydroxylation is in dynamic equilibrium at the $\{100\}_{\mathrm{Mn}_2\mathrm{O}_4}$ interface. The difference to the $\{100\}_\mathrm{Li}$ interface is that the equilibrium of surface PT reactions is shifted towards the occurrence of OH$^-$ ions at the $\{100\}_{\mathrm{Mn}_2\mathrm{O}_4}$ interface. Another frequent PT shown in Figure \ref{fig:PT_reactions} (c) happens in the perpendicular direction to the process in Figure \ref{fig:PT_reactions} (b) making the proton transport two-dimensional (Figure \ref{fig:PT_directions} (b)). Different dimensionalities of PT networks at different solid surfaces have been observed for ZnO-water interfaces before.\cite{Hellstroem2019} In the PT shown in Figure \ref{fig:PT_reactions} (c) the proton is transferred within the first liquid layer from an H$_2$O molecule to an OH$^-$ ion, i.e., adlayer PT (red).\cite{Quaranta2017} Further, the aforementioned intermediate H$_3$O$^+$ ions can alternatively transfer another proton to an adjacent H$_2$O molecule in the first liquid layer in the same direction as the PT in Figure \ref{fig:PT_reactions} (c) (yellow lines in Figure \ref{fig:PT_directions} (b)). In this way, long-range proton transport by PT reactions at the $\{100\}_{\mathrm{Mn}_2\mathrm{O}_4}$ interface is possible. The intermediate H$_3$O$^+$ ions also explain the relatively high concentration of H$_3$O$^+$ in Figure \ref{fig:oxygen_species} (b).

As for the $\{100\}$ interfaces, surface PTs are a large fraction of the PT reactions at the $\{110\}_{\mathrm{LiMnO}_2}$ interface (Figure \ref{fig:PT_species} (c)). The dominant PT reaction, where the proton of an H$_2$O molecule is transferred to the O$^{2-}$ ion of the first solid layer, can be followed by a PT from an OH$^-$ ion of the first solid layer to the just created OH$^-$ ion (Figure \ref{fig:PT_reactions} (d)). The OH$^-$ ion in the first liquid layer is therefore an intermediate species in this two-step transfer (violet zigzag lines in Figure \ref{fig:PT_directions} (c)). In contrast to the other interfaces, OH$^-$ ions in the first liquid layer can also transfer their proton to another OH$^-$ ion forming surface exposed O$^{2-}$ ions at the $\{110\}_{\mathrm{LiMnO}_2}$ solid surface, i.e., H$_2$O molecules are deprotonated twice (second closest violet peak to the interface in Figure \ref{fig:PT_species} (c) and vertical violet lines in Figure \ref{fig:PT_directions} (c)). In addition, adlayer PTs can happen as well yielding in total a two-dimensional PT network.

At the $\{110\}_{\mathrm{MnO}_2}$ interface PTs between H$_2$O molecules and OH$^-$ ions are dominant (Figure \ref{fig:PT_species} (d)). In contrast to the other interfaces, these reactions can also form H$_2$O molecules in the first solid layer. Hence, the O$^{2-}$ ions of the solid can be protonated twice. The protonation of an OH$^-$ ion in the first solid layer by an H$_2$O molecule from the first liquid layer is shown in Figure \ref{fig:PT_reactions} (e). This reaction can be an indication of solid surface reconstruction and dissolution at longer time scales as the boundary between solid and liquid changes. Surface PTs have been observed at the $\{110\}_{\mathrm{MnO}_2}$ interface as well. The PT pathways form again a two-dimensional zigzag pattern at the $\{110\}_{\mathrm{MnO}_2}$ interface (Figure \ref{fig:PT_directions} (d)). The PT networks at the $\{110\}$ interfaces are thus very different to those at the $\{100\}$ interfaces (Figures \ref{fig:PT_directions} (a) to (d)). 

\begin{figure*}[htb!]
\centering
\includegraphics[width=\textwidth]{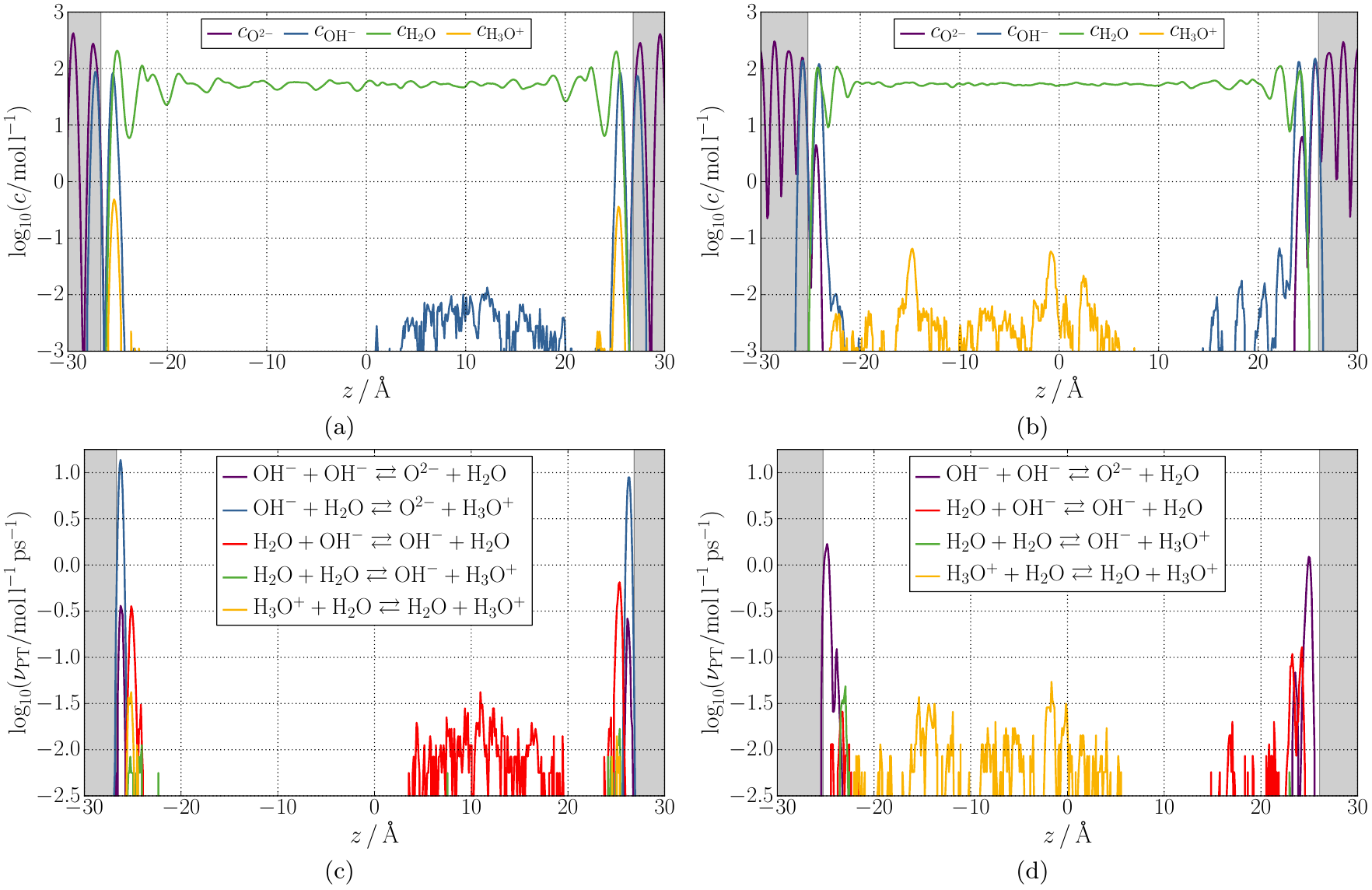}
\caption{Decadic logarithm of the concentration $c$ for different oxygen species as a function of the oxygen $z$ coordinate of the (a) $\{100\}_{\mathrm{Mn}_2\mathrm{O}_4}$ and (b) $\{110\}_{\mathrm{LiMnO}_2}$ Li$_x$Mn$_2$O$_4$-water interface systems and decadic logarithm of the PT reaction rates $\nu_\mathrm{PT}$ between different water species as a function of the mean $z$ coordinate of the transferred H atom for the (c) $\{100\}_{\mathrm{Mn}_2\mathrm{O}_4}$ and (d) $\{110\}_{\mathrm{LiMnO}_2}$ Li$_x$Mn$_2$O$_4$-water interface systems. The solid phase is highlighted by a gray background.}\label{fig:Grotthuss}
\end{figure*}

In conclusion, PT processes are much more frequent at the interface compared to the bulk liquid. Many different processes can happen which can be very specific to the interface structure. Larger interface areas in principle increase the PT reaction rates, but some reactions, as observed, e.g., for solvent-assisted PTs and the formation of surface exposed O$^{2-}$ ions, only happen at particular sites. Only an increase of these surface sites would lead to a higher activity highlighting the importance of the site-specific atomistic understanding. The reaction mechanisms at the different interfaces determine how fast long-range transport of reactive species by PT reactions can be. The strongly bound water structure of the interface promotes alternating forward and backward reactions. The rate itself does not seem to be an sufficient indicator of the transport. An increased number of possible paths can increase the effective transport of the protons. The highest activity is found between the first solid and first liquid layer followed by processes taking place within the first liquid layer. Consequently, the structure of the first liquid layer is also of major importance for the reactivity.

As shown already in Figure \ref{fig:oxygen_species} a very few OH$^-$ and H$_3$O$^+$ ions can escape from the interface. In two of the simulations this process was successful as shown in Figures \ref{fig:Grotthuss} (a) and (b). The OH$^-$ ion in Figure \ref{fig:Grotthuss} (c) and the H$_3$O$^+$ ion in Figure \ref{fig:Grotthuss} (d) diffuse via the Grotthuss mechanism through the bulk. The H$_3$O$^+$ thereby escaped from the left interface and diffused for about 265\,ps in the bulk until it recombined at the initial interface. The diffusion via the Grotthuss mechanism is very fast as all PT events in the bulk originate from a single H$_3$O$^+$ species present only for 5.3{\%} of the simulation time. The PT rate from H$_2$O to OH$^-$ divided by the OH$^-$ concentration in the case of the $\{100\}_{\mathrm{Mn}_2\mathrm{O}_4}$ system, i.e., the difference between the red graph in Figure \ref{fig:Grotthuss} (c) and the blue graph in Figure \ref{fig:Grotthuss} (a), shows that the OH$^-$ ion in the bulk undergoes more than one PT per picosecond. In the liquid layer closest to the interface the rate is below one per picosecond. Liquid layers further away from the interface can also reach as high reaction rates per particle as in the bulk liquid. Thus, the high concentration of reactive species at the interface is the reason for the high PT reaction rates. The very specific reaction paths present at the interface cannot take place in the bulk.

%%%%%%%%%%%%%%%%%%%%%%%%%%%%%%%%%%%%%%%%%%%%%%%%%%%%%%%%%%%%%%%%%%%%%%%%%%%%%%%%%%%%%%%%%%%%%%%%%%%%
\section{Conclusion}
%%%%%%%%%%%%%%%%%%%%%%%%%%%%%%%%%%%%%%%%%%%%%%%%%%%%%%%%%%%%%%%%%%%%%%%%%%%%%%%%%%%%%%%%%%%%%%%%%%%%

A high-dimensional neural network potential has been used to perform first principles-quality simulations of several Li$_x$Mn$_2$O$_4$ spinel-water interfaces. These simulations reveal spontaneous water dissociation and the formation of a hydroxide layer at several solid surface cuts. The different water species can arrange in strongly bound water layers with significantly reduced mobility at the interface, whose structures are highly dependent on the underlying solid surface. The oxygen atoms of the different water species in the first liquid layer tend to mimic the bulk oxygen coordination of Li and Mn ions because these binding sites are preferred by the solid. A second strongly bound water layer is formed at interfaces with a large fraction of hydroxide ions. Its structure deviates from the face centered cubic lattice of bulk oxygen to optimize the interaction with the bulk liquid.

A high-dimensional neural network for spin prediction allows to determine the influence of the solid-liquid interface on the Mn oxidation state distribution. Weak interaction with water seems to be an indicator for only little formation of Mn$^\mathrm{II}$ ions at the interface. An increasing number of coordinating oxide ions of interfacial Mn ions stabilizes higher Mn oxidation states. The mobility of e$_\mathrm{g}$ electrons is increased in the vicinity of the interface compared to the bulk. Consequently, higher battery performance can be achieved with smaller particles sizes while smaller particles are also more prone to degradation since soluble Mn$^\mathrm{II}$ ions are formed at the interface.

Proton transfer occurs between water molecules and oxide ions at the solid surface. The formed hydroxide ions originating from water molecules are stabilized in the vicinity of Mn ions. Active sites at the interface increase the amount of water dissociation significantly compared to the bulk liquid. Beyond the solid surface, the reactivity is also highly dependent on the structure of the water contact layer because a large part of the proton transfer reactions takes place within this layer. The mobility of the water species at the interface is reduced. However, the protons can be transported by PT reactions depending on the effective dimensionality of the surface-specific proton transfer network. The number of possible paths as well as the underlying mechanisms are highly dependent on the interface structure. Therefore, a larger interface area increases the reactivity only if this interface includes the respective active sites.

In summary, in the present work a first step has been taken towards an understanding of site-specific reactions at the atomic scale to obtain control tactics for higher performance and durability of battery materials as well as higher reactivity of catalysts.

\section*{Supplementary Material}

See Supplementary Material for (I.A.) construction of the reference data set,\cite{Artrith2012, Eckhoff2019, Eckhoff2020a, Eckhoff2020b, Eckhoff2021} (I.B.) determination of the atom-centered symmetry function parameters,\cite{Behler2011} (I.C.) RuNNer settings,\cite{Eckhoff2020a, Eckhoff2020b} (II.A.) Jahn-Teller distortions,\cite{Eckhoff2020a} (II.B.) charge conservation,\cite{Zurbenko1986, Eckhoff2020b} (II.C.) hydrogen bonds, (II.D.) structure of the interface, (II.E.) residence correlation function, and (II.F.) proton transfer reaction rates.

\begin{acknowledgments}
This project was funded by the Deutsche Forschungsgemeinschaft (DFG, German Research Foundation) - 217133147/SFB 1073, project C03. We gratefully acknowledge computing time provided by the Paderborn Center for Parallel Computing (PC$^2$) and by the DFG project INST186/1294-1 FUGG (Project No.\ 405832858). Discussions with Peter E. Bl\"ochl are gratefully acknowledged.
\end{acknowledgments}

\section*{Data Availability}

The data that support the findings of this study are available from the corresponding author upon reasonable request.

\section*{Conflicts of Interest}

The authors declare no conflicts of interest.

\section*{References}

\bibliography{bibliography}

\end{document}